\documentclass[twocolumn]{autart}   

\usepackage{graphicx}       
\usepackage{amsmath}
\usepackage{amssymb}
\usepackage{siunitx}
\usepackage{enumitem}
\usepackage[dvipsnames]{xcolor}
\usepackage{booktabs}
\usepackage[hidelinks]{hyperref}
\usepackage[round]{natbib}

\newcommand{\ini}{\alpha}
\newcommand{\fin}{\omega}
\newcommand{\Ri}[1]{\mathbb{R}^{#1}}
\newcommand{\Rip}{\mathbb{R}_{\ge 0}}
\newcommand{\cont}[1]{\mathcal{C}^{#1}}
\newcommand{\I}[1]{{{I}}_{#1}}
\newcommand{\0}[1]{{{0}}_{#1}}
\newcommand{\Mats}{\mathbb{M}}
\newcommand{\bigO}{O}
\DeclareMathOperator*{\argmin}{arg\,min}
\DeclareMathOperator*{\sat}{sat}
\DeclareMathOperator*{\distance}{dist}
\newcommand{\inv}{^{-1}}
\newcommand{\pinv}{^{\dagger}}
\newcommand{\transp}{^{\mathsf{T}}}

\newcommand{\colvec}{\text{col}}
\newcommand{\jac}[1]{D{#1}}
\newcommand{\hess}[1]{D^2{#1}}
\DeclareMathOperator{\closure}{cl}
\DeclareMathOperator{\interior}{int}

\newcommand{\nom}{\star}
\newcommand{\nomorb}{\mathcal{O}}
\newcommand{\maneuver}{\mathcal{M}}
\newcommand{\nomflow}{\mathcal{F}}
\newcommand{\tubespace}{\mathcal{T}}
\newcommand{\neighb}{\mathcal{N}}

\newcommand{\state}{{{x}}}
\newcommand{\Dstate}{{\dot{{x}}}}
\newcommand{\ffunc}{{{f}}}
\newcommand{\Efunc}{{{A}}}
\newcommand{\gfunc}{{{B}}}
\newcommand{\gfunccol}{{{b}}}
\newcommand{\gc}{{{q}}} 
\newcommand{\ac}{{{u}}}
\newcommand{\nfb}{k}
\newcommand{\lfb}{K}
\newcommand{\vc}{{{\Phi}}}
\newcommand{\vcs}{{{\phi}}}
\newcommand{\mg}{s}
\newcommand{\mgspace}{\mathcal{S}}
\newcommand{\nvel}{\rho}
\newcommand{\balpha}{\mathfrak{A}}
\newcommand{\bbeta}{\mathfrak{B}}
\newcommand{\bgamma}{\mathfrak{G}}
\newcommand{\tvc}{{e}}
\newcommand{\tvcjac}{\mathcal{E}_{\perp}}
\newcommand{\prj}{p}
\newcommand{\xp}{\state_{\prj}}
\newcommand{\prjspace}{\mathfrak{X}}
\newcommand{\prjspacesub}{\hat{\mathfrak{X}}}
\newcommand{\projjac}{\mathcal{P}}

\begin{document}

\begin{frontmatter}
\title{Orbital Stabilization of Point-to-Point Maneuvers \\ in Underactuated Mechanical Systems}

\thanks[footnoteinfo]{Corresponding author. 
This research was supported by the Research Council of Norway, grant  {262363}.
}

\author[ITK]{Christian Fredrik Sætre\thanksref{footnoteinfo}}\ead{christian.f.satre@ntnu.no}  ,   
\author[ITK]{Anton S. Shiriaev}\ead{anton.shiriaev@ntnu.no}               

\address[ITK]{Department  of Engineering Cybernetics, NTNU, O. S. Bragstads Plass 2D, 7034 Trondheim, Norway.} 


\begin{abstract}                
The task of inducing, via continuous static state-feedback control, an asymptotically stable heteroclinic orbit in a nonlinear control system is considered in this paper. The main motivation comes from the problem of ensuring convergence to a so-called point-to-point maneuver in an underactuated mechanical system. Namely, to a smooth curve in its state--control space which is consistent with the system dynamics and connects two (linearly) stabilizable equilibrium points. The proposed method uses a particular parameterization, together with a state projection onto the maneuver as to combine two linearization techniques for this purpose: the Jacobian linearization at the equilibria on the boundaries and a transverse linearization along the orbit. This allows for the computation of stabilizing control gains offline by solving a semidefinite programming problem. The resulting nonlinear controller, which simultaneously asymptotically stabilizes both the orbit and the final equilibrium, is time-invariant, locally Lipschitz continuous, requires no switching, and has a familiar feedforward plus feedback--like structure. The method is also complemented by synchronization function--based arguments for planning such maneuvers for mechanical systems with one degree of underactuation. Numerical simulations of the non-prehensile manipulation task of a ball rolling between two points upon the ``butterfly'' robot  demonstrates the efficacy of the synthesis.
\end{abstract}

\end{frontmatter}

\section{Introduction}
A \emph{point-to-point} (PtP) motion is perhaps the most fundamental of all motions in robotics: Starting from rest at a certain configuration (point), the task is to steer the system to rest at a different goal configuration. Often it can  also be beneficial, or even necessary, to know a specific predetermined motion which smoothly connects the two configurations, in the form of a curve in the state--control space which is consistent with the system dynamics---a \emph{maneuver} \citep{hauser1995maneuver}.  For instance, this ensures that the controls remain within the admissible range  along the nominal motion, and that neither any kinematic- nor dynamic constraints are violated along it. Knowledge of a maneuver is also especially important for an \emph{underactuated mechanical system} (UMS) \citep{spong1998underactuated,liu2013survey}.  Indeed, as an UMS has fewer independent controls (actuators) than degrees of freedom, any feasible motion must necessarily comply with the dynamic constraints which arise due to the system's underactuation \citep{shiriaev2005constructive}.

Planning such (open-loop) PtP maneuvers in an UMS, e.g. a swing-up motion of a pendulum-type system with several passive degrees of freedom, is of course  a nontrivial task in itself.
Suppose, however, that such a maneuver has been found. Then the next step is to design a stabilizing feedback  for it.
For non-feedback-linearizable systems (i.e. the vast majority)  this is also a nontrivial task. The challenge again lies in the lack of actuation, which may severely limit the possible actions the controller can take. This can make reference tracking controllers less suited for this purpose, as they, often unnecessarily so, are tasked with tracking one specific trajectory (among infinitely many) along the maneuver. 

For tasks  which do not require a specific \emph{timing} of the motion, one can instead design an \emph{orbitally stabilizing feedback}: a time-invariant state-feedback controller which (asymptotically) stabilizes the set of all the states along the maneuver---its \emph{orbit}. For a PtP maneuver, such a feedback controller is therefore equivalent to inducing an asymptotically stable \emph{heteroclinic orbit} in the resulting autonomous closed-loop system. Namely, an  invariant, one-dimensional manifold which (smoothly) connects the initial and final equilibrium points. There are some clear advantageous to such an approach: First,  all solutions initialized upon the  orbit asymptotically converge to the final equilibrium along the maneuver, with the behavior when evolving along it known a priori. Second,  by invoking a reduction principle \citep{el2013reduction},   the final equilibrium's asymptotic stability  is ensured by the orbit's  asymptotic stability. Third, the closed-loop system is time invariant.

In regard to the problem of  designing  such feedback, the maneuver regulation approach proposed in \citep{hauser1995maneuver} is of particular interest. There, the task of stabilizing---via static state feedback---non-vanishing (i.e. equilibrium-free) orbits of feedback-linearizable systems was considered; with the approach later extended to a class of non-minimum phase systems in normal form in \citep{al2002tracking}.  The key idea in these papers is to convert a linear tracking controller into a controller  stabilizing the orbit of a known maneuver. This is achieved by using a  projection of the system states onto the maneuver, a \emph{projection operator} as we will refer to it here, to recover the corresponding ``time'' to be used in the controller, thus eliminating its time dependence. The former tracking error therefore instead becomes a \emph{transverse} error---a weighted measure of the  distance from the current state to the maneuver's orbit.

It has long been known  for non-trivial orbits (e.g.periodic ones) that strict contraction in the directions transverse to it is equivalent to its asymptotic stability \citep{borg1960condition,hartman1962global,urabe1967nonlinear,hauser1994converse,zubov1999theory,manchester2014transverse}. Moreover, this contraction can be determined from a specific linearization of the system dynamics along the nominal orbit \citep{leonov1995local}, a so-called \emph{transverse linearization} \citep{hauser1994converse,shiriaev2010transverse,manchester2011transverse,saetre2020IFACWC}. Since this contraction occurs on transverse hypersurfaces, only the linearization of a set of transverse coordinates of dimension one less than the dimension of the state space needs to be stabilized; a fact which has been readily used to stabilize periodic orbits in UMSs  \citep{shiriaev2010transverse,surov2015case}.

For the purpose we consider in this paper, namely the design of a continuous (orbitally) stabilizing feedback controller for PtP motions with a known maneuver, one must also take into consideration the equilibria located at the boundaries of the motion. On the one hand, this directly excludes regular transverse coordinates--based methods  such as \citep{shiriaev2005constructive,shiriaev2010transverse,manchester2011transverse}, which would then require some form of control switching and/or orbit jumping {à la} those in \citep{la2009new,sellami2020ros}. The ideas proposed by \cite{hauser1995maneuver} in regard to maneuver regulation, on the other hand, can be modified as to also handle the equilibria, but suffers from other shortcomings: 1) the choice of projection operator is strictly determined by the tracking controller, thus excluding simpler operators, e.g., operators only depending on the configuration variables; while most importantly, 2) the requirement of a feedback-linearizable system and  constant feedback gains greatly limits its applicability to stabilize (not necessarily PtP) motions of both UMSs and nonlinear dynamical systems in general.

\textbf{Contributions.} 
 The main contribution of this paper is an approach that extends the applicability of the ideas in \citep{hauser1995maneuver} to a larger class of dynamical systems, as well as to different types of behaviors, including point-to-point (PtP) maneuvers.   The main novelty in our approach lies in the use of a specific parameterization of the maneuver, together with an operator providing a projection onto it. Roughly speaking,  this allows us to merge the transverse linearization with the regular Jacobian linearization at the boundary equilibria. This, in turn, allows us to derive a (locally Lipschitz) Lyapunov function candidate for the nominal orbit as a whole.
Specifically, the paper's main contributions are:
\begin{enumerate}
    \item Sufficient conditions ensuring that a (locally Lipschitz continuous) feedback controller orbitally stabilizes a known PtP maneuver of a nonlinear control-affine system; see Theorem~\ref{theorem:MR} in Section~\ref{sec:MainSection}.
    \item A constructive procedure allowing for the design  of such feedback by solving a semidefinite programming problem; see Proposition~\ref{prop:LMIcontDesign} in Section~\ref{sec:MainSection}.
    \item A synchronization function--based method for planning PtP maneuvers for a class of underactuated mechanical systems; see Theorem~\ref{theorem:PtPcondsUAS} in Section~\ref{sec:UnAcSys}.
    \item Arguments facilitating the generation of orbitally stable PtP motions of a ball rolling between any two points upon the frame of the ``butterfly'' robot; see Proposition~\ref{prop:BRprop} in Section~\ref{sec:NPexampleSection}
\end{enumerate}
Note also that, with only minor modifications, these statements can also be used to generate and orbitally stabilize (hybrid) periodic motions, or to ensure contraction toward a  non-vanishing motion defined on a finite time interval.

All proofs are given in the Appendix. A statement is ended by $\square$ if its proof is not provided.

\textbf{Notation.}
$\I{n}$ denotes the $n\times n$ identity matrix and  $\0{n\times m}$ an $n\times m$ matrix of zeros, with $\0{n}=\0{n\times n}$. For  $\mgspace\subset\Ri{n}$,  $\interior{(\mgspace)}$ denotes its interior
and $\closure(\mgspace)$ its closure. For  $\state\in\Ri{n}$, $\|\state\|=\sqrt{\state\transp x}$.  For some $\epsilon>0$ and $\state\in\Ri{n}$ we denote $\mathcal{B}_{\epsilon}(\state):=\{y\in\Ri{n}: \ \|\state-y\|<\epsilon\}$. For  column vectors $x$ and $ y$,  $\colvec(x,y):=[x\transp,y\transp]\transp$ is used. For $x,y\in\Ri{n}$ we denote $\mathfrak{L}(x,y)=\{x+(y-x)\iota, \iota\in[0,1]\}$. If $h:\Ri{n}\to\Ri{m}$ is $\cont{1}$, then $\jac{h}:\Ri{n}\to\Ri{m\times n}$  denotes its Jacobian matrix, and if $m=1$ then $\hess{h}:\Ri{n}\to\Ri{n\times n}$ denotes its Hessian matrix. If $\mg\mapsto h(\mg)$ is differentiable at $\mg\in\mgspace\subseteq\Ri{}$, then  $h'(\mg)=\frac{d}{d\mg}h(\mg)$.  $\|\sigma(x)\|=\bigO(\|x\|^{k})$ if there exists $c>0$ such that $\|\sigma(x)\|\le c \|x\|^k$ as $\|x\|\to 0$.  $\mathbb{M}^n_{\succ 0}$ (resp. $\mathbb{M}^n_{\succeq 0}$)  denotes the set of all  real, symmetric, positive (resp. semi-) definite $n\times n$ matrices,  such that $R\succ \0{n}$ if $R\in\mathbb{M}^n_{\succ 0}$.

\section{Problem formulation}\label{sec:ProblemFormulation}
Consider a nonlinear control-affine system
\begin{equation}\label{eq:DynSys}
    \Dstate=\ffunc(\state)+\gfunc(\state)\ac
\end{equation}
with state $\state\in\Ri{n}$ and with ($m\le n$) controls $\ac\in\Ri{m}$.  It is assumed that both $\ffunc:\Ri{n}\to\Ri{n}$ and the columns of the full-rank matrix function $\gfunc:\Ri{n}\to\Ri{n\times m}$, denoted $\gfunccol_i(\cdot)$,  are twice continuously differentiable ($\cont{2}$).

Let the pair $(\state_e,\ac_e)\in\Ri{n}\times\Ri{m}$ correspond to an equilibrium  of  \eqref{eq:DynSys}, i.e., $\ffunc(\state_e)+\gfunc(\state_e)\ac_e\equiv\0{n\times 1}$. If we denote 
\begin{equation}\label{eq:Edef}
    {\Efunc}(\state,\ac):=\jac{\ffunc}(\state)+\sum_{i=1}^m \jac{ \gfunccol_i}(\state)\ac_i,
\end{equation}
 then the  (forced) equilibrium point, $\state_e$, is said to be linearly stabilizable   if there exists some $K\in\Ri{m\times n}$ such that  $\Efunc({\state}_e,{\ac}_e)+\gfunc({\state}_e)K$  is Hurwitz (stable). That is,  the full-state feedback $\ac=\ac_e+K(\state-\state_e)$ then renders $\state_e$ an exponentially stable equilibrium of \eqref{eq:DynSys}.

We will assume knowledge of a point-to-point (PtP) maneuver  connecting  two separate  linearly-stabilizable equilibrium points of \eqref{eq:DynSys}. Specifically, we assume that   a so-called $\mg$-parameterization of the maneuver is known:
\begin{defn}\label{def:s-param_mamever}
Let $(\state_{\ini},\ac_{\ini})$ and $(\state_{\fin},\ac_{\fin})$, $\state_{\ini}\neq\state_{\fin}$, be  linearly-stabilizable equilibrium points of \eqref{eq:DynSys}.  For  $\mgspace:=[\mg_{\ini},\mg_{\fin}]\subset\Ri{}$, $\mg_{\ini}<\mg_{\fin}$, the   triplet of functions
\begin{equation}\label{eq:maneuverTriplet}
    \state_\nom:\mgspace\to\Ri{n}, \ \  \ac_\nom:\mgspace\to\Ri{m}, \ \ \text{and} \ \ \nvel:\mgspace\to\Rip{},
\end{equation}
    constitute an  $\boldsymbol{\mg}$\textbf{-parameterization} of the PtP  maneuver
\begin{equation*}
    \maneuver:=\{(\state,\ac)\in\Ri{n}\times\Ri{m}: \ \state=\state_\nom(\mg), \ \ac=\ac_\nom(\mg), \ \mg\in\mgspace\}
\end{equation*}
   of \eqref{eq:DynSys},  whose boundaries are $(\state_{\ini},\ac_{\ini})$ and $(\state_{\fin},\ac_{\fin})$, 
 if 
\begin{itemize}
\item[\textbf{P1}] $\state_\nom(\cdot)$ is of class $\cont{2}$ and traces out a non-self-intersecting curve, while  $\ac_\nom(\cdot)$ and $\nvel(\cdot)$ are $\cont{1}$;\footnote{While the notion of an $\mg-parameterization$ can be relaxed in regard to the smoothness of the triplet $(\state_\nom,\ac_\nom,\nvel)$, we will, for simplicity's sake,  require that \textbf{P1} holds in this paper.  } 
\item[\textbf{P2}] $(\state_\nom(\mg_i),\ac_\nom(\mg_i))=(\state_i,\ac_i)$ \text{for both} 
    $i\in\{\ini,\fin\}$;
\item[\textbf{P3}]  $\nvel(\mg_{\ini})=\nvel(\mg_{\fin})\equiv 0$, while $\nvel(\mg)>0$   for all $\mg\in\interior{\mgspace}$;
 \item[\textbf{P4}] $\|\nomflow(\mg)\|>0$ for all $\mg\in\mgspace$,  \text{where} $ \nomflow(\mg):=\state_\nom'(\mg)$; 
\item[\textbf{P5}] $\nomflow(\mg)\nvel(\mg)=\ffunc(\state_\nom(\mg))+\gfunc(\state_\nom(\mg))\ac_\nom(\mg)$  for all $\mg\in\mgspace$.
\end{itemize}
\end{defn}

The definition requires some further comments. Given an $\mg$-parameterized maneuver $\maneuver$,  we denote   by
\begin{equation}\label{eq:nomOrb}
    \nomorb:=\{\state\in\Ri{n}: \ \state=\state_\nom(\mg), \quad \mg\in\mgspace\}
\end{equation}
 its corresponding \emph{orbit} (i.e. its projection upon state space). Due to the properties of $\maneuver$ stated in Definition~\ref{def:s-param_mamever}, one may in fact consider  $\nomorb$ to consist of a (forced) \emph{heteroclinic orbit} of \eqref{eq:DynSys} and its limit points:  by \textbf{P1},  $\nomorb$ is  a $\cont{2}$-smooth, one-dimensional embedded submanifold of $\Ri{n}$; by \textbf{P2}, its boundaries correspond to two separate (forced) equilibrium points of \eqref{eq:DynSys}; whereas by \textbf{P3}, \textbf{P4} and \textbf{P5}, it  is a  controlled invariant set of \eqref{eq:DynSys} that contains no (forced) equilibrium points on its interior.

Here the latter point can be  verified by viewing the curve parameter $\mg=\mg(t)$ as a solution to 
\begin{equation}\label{eq:nvel}
    \dot{\mg}=\nvel(\mg).
\end{equation}
Since $\dot{\state}_\nom(\mg(t))=\state_\nom'(\mg(t))\dot{\mg}(t)=\nomflow(\mg(t))\nvel(\mg(t))$ by the chain rule, one finds, by inserting this into the left-hand side of  the expression in  \textbf{P5}, that $\maneuver$ is consistent with the dynamics \eqref{eq:DynSys}. Thus, whereas $\|\dot{\state}_\nom(\mg(t))\|\equiv 0$ for $\mg(t)\in\{\mg_{\ini},\mg_{\fin}\}$, the key aspect of  an $\mg$-parameterization is that the regularity condition \textbf{P4} holds for $\state_\nom(\cdot)$, as $\nvel(\cdot)$ instead   vanishes at the boundaries.\footnote{As $\nvel(\cdot)$ is required to be  $\cont{1}$ and $\nvel(\mg_{\ini})=\nvel(\mg_{\fin})\equiv 0$, the rate at which $\mg(\cdot)$  convergence to $\mg_{\fin}$ (resp. $\mg_{\ini}$) in positive (resp. negative) time from within $\mgspace$ can be at most exponential, which  corresponds to $\nvel'(\mg_{\fin})<0$ (resp. $\nvel'(\mg_{\ini})>0$). } This property allows for a compact representation of the motion, something which is clearly seen from  the nominal state curve's arc length:
 $\int_{-\infty}^{\infty} \lVert \dot{\state}_\nom(\mg(\tau))\rVert d\tau =\int_{\mg_{\ini}}^{\mg_{\fin}}\lVert \nomflow(\sigma)\rVert d\sigma$. It is also vital to the approach we suggest, as it  allows one to  construct a well-defined projection onto the maneuver.

 For $\nomorb$ as defined in \eqref{eq:nomOrb}, denote $\distance(\nomorb,\state):=\inf_{y\in\nomorb}\|\state-y\|$. We aim to solve the following problem  in this paper:
\begin{prob}(Orbital Stabilization)
\label{probl:EOSofPPm}
For \eqref{eq:DynSys}, construct a control law $\ac=\nfb(\state)$, with $\nfb:\Ri{n}\to\Ri{m}$ locally Lipschitz in a neighborhood of  $\nomorb$ and satisfying  $\nfb(\state_\nom(\mg))\equiv \ac_\nom(\mg)$ for all $\mg\in\mgspace$,  such that  $\nomorb$ is an asymptotically stable set  of the closed-loop system. Namely, for every $\epsilon>0$,  there is a $\delta>0$, such that for any solution $\state(\cdot)$  of the closed-loop system  satisfying $\distance(\nomorb,\state(t_0))<\delta$,  it  is implied that  $\distance(\nomorb,\state(t))<\epsilon$ for all  $t\ge t_0$ \emph{(stability)}, and that  $\distance(\nomorb,\state(t))\to 0$  as $t\to \infty$ \emph{(attractivity)}. 
\end{prob}

Note that the asymptotic stability of $\nomorb$  is equivalent to the \emph{asymptotic orbital stability}   of all the solutions upon it \citep{hahn1967stability,leonov1995local,urabe1967nonlinear,zubov1999theory}. 
Thus Problem~\ref{probl:EOSofPPm} is a so-called \emph{orbital stabilization problem},  which can be stated for any type of orbit (equilibrium points, (hybrid) periodic orbits, etc.). Moreover, as we here consider a heteroclinic orbit on which all solutions converge to $\state_{\fin}$, a solution to Problem~\ref{probl:EOSofPPm} also  implies the  (local) asymptotic stability of $\state_{\fin}$  by the reduction principle in \citep{el2013reduction}.

\section{Preliminaries}\label{sec:Preliminaries}
\subsection{Projection operators}\label{sec:ProjOp}
A key part of our approach is a projection onto
the set $\nomorb$ defined in \eqref{eq:nomOrb}.
We define such \emph{projection operators} in terms of a specific $\mg$-parameterization (see Def.~\ref{def:s-param_mamever}) next. 
\begin{defn}(Projection operators for PtP maneuvers)
\label{def:ProjOp}
Let $\prjspace\subset \Ri{n}$ denote a simply-connected neighborhood of $\nomorb$, whose interior can be partitioned into three subsets, denoted $\mathcal{H}_{\ini}$, $\tubespace$ and $\mathcal{H}_{\fin}$  (i.e., $\closure(\prjspace)=\closure(\mathcal{H}_{\ini}\cup\tubespace\cup\mathcal{H}_{\fin})$), which are such that
\begin{itemize}[leftmargin=5mm]
    \item  $\tubespace$ is a  tubular neighborhood  of $\nomorb$;
    \item $\mathcal{B}_{\epsilon}(\state_{\ini})\backslash \tubespace\subset\closure{(\mathcal{H}_{\ini})}$ and $\mathcal{B}_{\epsilon}(\state_{\fin})\backslash \tubespace\subset\closure{(\mathcal{H}_{\fin})}$, for some   $\epsilon>0$;
    \item $\closure(\tubespace)\cap\closure(\mathcal{H}_{i})\neq \emptyset$ $\forall i\in\{\ini,\fin\}$,  $\closure(\mathcal{H}_\ini)\cap\closure(\mathcal{H}_{\fin})=\emptyset$.
\end{itemize}
A map $ \prj:\prjspace\to\mgspace$ is said be to  a  \textbf{projection operator} for $\nomorb$ if it  Lipschitz continuous and well defined within its domain $\prjspace$, as well as satisfies
\begin{enumerate}
\item[\textbf{C1}] $\prj(\state_\nom(\mg))\equiv\mg$  for all $  \mg\in\mgspace$;
 \item[\textbf{C2}] $\prj(\mathcal{H}_{\ini})\equiv \mg_{\ini}$, $\prj(\mathcal{H}_{\fin})\equiv \mg_{\fin}$, and $\prj(\interior{(\tubespace)})\in\interior{(\mgspace)}$;
 \item[\textbf{C3}] $\prj(\cdot)$ is $\cont{r}$, $r\ge 2$, within  $\mathcal{H}_{\ini}$, $\tubespace$ and $\mathcal{H}_{\fin}$.\footnote{While one can generally relax the condition in \textbf{C3} to $r \ge 1$, we require $r\ge 2$ for the approach we suggest in this paper.}   
\end{enumerate}
\end{defn}

\begin{figure}
    \centering
    \includegraphics[width=\linewidth ]{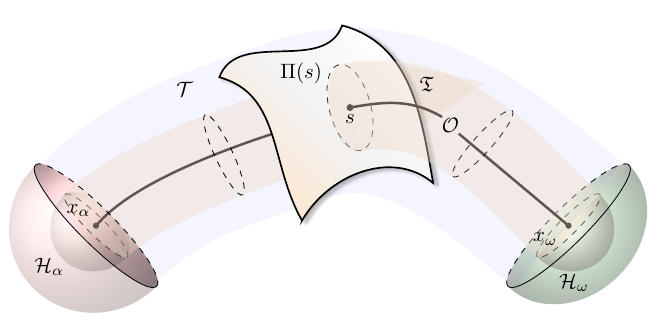}
    \caption{Illustration of the moving Poincaré section $\Pi({\mg})$, defined in \eqref{eq:MPS}, ``traveling'' along the orbit $\nomorb$  whose boundaries are $\state_{\ini}$ and  $\state_{\fin}$. The gradient of the projection operator is assumed to be nonzero and well defined within the blue-shaded tubular neighborhood $\tubespace$. Within the darkly shaded hemispheres $\mathcal{H}_{\ini}$ and $\mathcal{H}_{\fin}$, on the other hand, the gradient vanishes as the projection operator projects the states onto the respective equilibrium therein. The aim of this paper is to guarantee the existence of a positively invariant neighborhood $\mathfrak{T}$, within which all solutions converge to $\nomorb$.} 
    \label{fig:MPS}
\end{figure}

In order to provide some intuition behind the need for the conditions stated in Definition~\ref{def:ProjOp}, we define the set
 \begin{equation}\label{eq:MPS}
    \Pi({\mg}):=\{\state\in \prjspace: \prj(\state)={\mg}\}.    
 \end{equation}
As is illustrated in Figure~\ref{fig:MPS},  for some ${\mg}\in\interior{(\mgspace)}$, this set traces out a hypersurface, a so-called  \emph{moving Poincaré section} \citep{leonov2006generalization,shiriaev2010transverse}, whose tangent space at $\state_\nom(\mg)$ is  orthogonal to the transpose of
\begin{equation}\label{eq:ProjOpJac}
        \projjac(\mg):=\jac{\prj}(\state_\nom(\mg)).
\end{equation}
 By Condition \textbf{C1},  it follows that $\projjac(\mg)\nomflow(\mg)\equiv 1$ for all $\mg\in\mgspace$, from which,  in turn, one can deduce that the surface $\Pi({\mg})$ is locally transverse to $\nomflow({\mg})$. The tubular neighborhood $\tubespace$ in the definition (consider the blue-shaded tube in Figure~\ref{fig:MPS}) is therefore guaranteed to everywhere have a nonzero radius as $\nomorb$ does not have any self-intersections (see \textbf{P1} in Def.~\ref{def:s-param_mamever}). It can be taken as any connected subset of $\bigcup_{\mg\in\interior{\mgspace}}\Pi(\mg)$ such that the surfaces $\Pi(\mg_1)\cap\tubespace$ and $\Pi(\mg_2)\cap\tubespace$ are locally disjoint for any $\mg_1,\mg_2\in\interior{(\mgspace)}$, $\mg_1\neq \mg_2$.  Thus   $\jac{\prj}(\state)$  is nonzero, bounded and of class $\cont{r-1}$ for any  $\state$ within $\tubespace$.

Conditions \textbf{C2} and \textbf{C3}, on the other hand, guarantee the existence of the two open half-ball-like regions, $\mathcal{H}_{\ini}$ and $\mathcal{H}_{\fin}$, contained in  $\Pi(\mg_{\ini})$ and $\Pi(\mg_{\fin})$, respectively (see the darkly shaded semi-ellipsoids in Figure~\ref{fig:MPS}). As a consequence, $\|\jac{\prj}(\state)\|\equiv 0$ for all $\state \in {\mathcal{H}}_{\ini}\cup {\mathcal{H}}_{\fin}$, and hence  $\prj(\cdot)$ is $\cont{2}$ almost everywhere  within $\prjspace$, except at  $\prjspace_{\ini}:=\lim_{\mg\to\mg_{\ini}^+}\Pi(\mg)$ and $\prjspace_{\fin}:=\lim_{\mg\to\mg_{\fin}^-}\Pi(\mg)$, which correspond to the intersections of the boundary of $\tubespace$ with the boundaries of $\mathcal{H}_{\ini}$ and $\mathcal{H}_{\fin}$, respectively.

The following statements shows  that one can obtain projection operators  satisfying Definition~\ref{def:ProjOp}, which are similar to those in \cite{hauser1995maneuver}.

\begin{prop}\label{prop:GenProjOp}
Given a PtP maneuver as by Definition~\ref{def:s-param_mamever}, let the smooth matrix-valued function $\Lambda:\mgspace\to\mathbb{M}^n_{\succeq 0}$ be such that   $\mathfrak{h}(\mg):=\Lambda(\mg)\nomflow(\mg)$ is of class $\cont{2}$ on $\mgspace$, and  $\nomflow\transp(\mg)\Lambda(\mg)\nomflow(\mg)>0$ holds for all $\mg\in\mgspace$ . Then there is  an $\epsilon>0$ and a neighborhood $\prjspace$ of $\nomorb$, such that
    \begin{equation}\label{eq:GenProjOp}
    \prj(\state)=\argmin_{\substack{\mg\in\mgspace \\ 
    \mathfrak{L}(\state_\nom(\mg),\state)\subset\mathcal{B}_\epsilon(\state_\nom(\mg))
    }
     }
    \left[(\state-\state_\nom(\mg))\transp\Lambda(\mg)(\state-\state_\nom(\mg))\right]
\end{equation}
 is a projection operator for  $\nomorb$ (see Def.~\ref{def:ProjOp}), with $\mathcal{B}_\epsilon(\state_\nom(\mg))\subset\prjspace$  for all $\mg\in\mgspace$. Moreover, 
 \begin{equation}\label{eq:genProjOpJac}
     \projjac(\mg):=\jac{\prj}(\state_\nom(\mg))=\frac{\nomflow\transp(\mg)\Lambda(\mg)}{\nomflow\transp(\mg)\Lambda(\mg)\nomflow(\mg)}
 \end{equation}
 holds for its Jacobian matrix $\jac{\prj}(\cdot)$  evaluated inside  $\tubespace$.
 \end{prop}

Note  that, in order to effectively compute such operators,   knowledge of the hypersurfaces $\prjspace_\ini$ and $\prjspace_\fin$  can be used to locally partition $\prjspace$ into its respective subsets (see Def.~\ref{def:ProjOp}), with \eqref{eq:GenProjOp}  then generally having to be solved numerically when $\state$ is in $\tubespace$ (see also \citep{hauser1995maneuver}).
Notice also that
   $\Lambda(\cdot)$ is not required to be positive definite nor constant; indeed, for certain maneuvers, this  may allow one to use  operators depending  only on a few state variables and which can be directly evaluated rather than found numerically  (cf. Ex.~\ref{example:DI} and Sec.~\ref{sec:SimRes}). 

\subsection{Implicit representation of the orbit}\label{sec:ProjBasedCoords}
Given a projection operator  ${\prj}:\prjspace\to\mgspace$ as by Definition \ref{def:ProjOp}, we denote by  $\xp(\state):=(\state_\nom\circ\prj)(\state)$ the corresponding projection onto $\nomorb$, and define  the following function:
\begin{equation}\label{eq:eTC}
    \tvc(\state):=\state-\xp(\state). 
\end{equation}
From the properties of $\state_\nom(\cdot)$ and $\prj(\cdot)$ (see Def.~\ref{def:s-param_mamever} and Def.~\ref{def:ProjOp}, respectively), it follows that $\tvc=\tvc(\state)$ is well defined for $\state\in\prjspace$,  locally Lipschitz in a neighborhood of $\nomorb$, and therefore twice continuously differentiable  everywhere therein except at the two hypersurfaces $\prjspace_{\ini}$ and $\prjspace_{\fin}$ on the orbit's boundaries.  Most importantly, however,
is the fact that the zero-level set of this function corresponds to the nominal orbit $\nomorb$ which we aim to stabilize, while, locally, its magnitude is nonzero away from it. Our goal will therefore be to design a control law which guarantees the existence of a positively invariant  neighborhood $\mathfrak{T}$  of $\nomorb$  (see Fig.~\ref{fig:MPS}) within which $\tvc$  converges to zero. 

With this goal in mind,  observe from the definition of a  projection operator  (Def. ~\ref{def:ProjOp}) that one  may   interpret   $\tvc(\state)$ differently depending on where in $\prjspace$ the current state is located.
Indeed,   consider  the open sets $(\mathcal{H}_{\ini},\mathcal{H}_{\fin})$ and the tube $\tubespace$ introduced in Section~\ref{sec:ProjOp}.  
Clearly, whenever $\state\in{\mathcal{H}}_i$ for a fixed $i\in\{\ini,\fin\}$,  one  has $\tvc=\state-\state_i$ as  $\prj(\state)\equiv\mg_i$, and thus  $\jac{\tvc}(\state)=\I{n}$ therein.
For  $\state\in\tubespace$, on the other hand, the function $\tvc(\cdot)$  forms an excessive set of so-called \emph{transverse coordinates}  \citep{saetre2020IFACWC}. This can be observed from its Jacobian matrix evaluated along the orbit, which inside of $\tubespace$  is  given by
\begin{equation}\label{eq:tvcjac}
    \tvcjac(\mg):=\jac{\tvc}(\state_\nom(\mg))=\I{n}-\nomflow(\mg)\projjac(\mg)
\end{equation}
with $\projjac$ defined in \eqref{eq:ProjOpJac}. Since $\projjac(\mg)\nomflow(\mg)\equiv 1$, the matrix $\tvcjac(\mg)$ can be used to project any vector $x\in\Ri{n}$ upon the hyperplane orthogonal to $\projjac\transp(\mg)$.  As it will appear throughout this paper, we recall some of its  properties:
\begin{lem}[\cite{saetre2020IFACWC}]\label{lemma:tvcjac}
For all $\mg\in\mgspace$, the matrix function $\tvcjac:\mgspace\to\Ri{n\times n}$ defined in \eqref{eq:tvcjac} is a projection matrix, i.e. $\tvcjac^2(\mg)=\tvcjac(\mg)$; its rank is  $n-1$; while $\projjac(\mg)$ and $\nomflow(\mg)$ span its left- and right annihilator spaces, respectively. \qed 
\end{lem}

\subsection{Merging two types of linearizations }\label{sec:mergingLins}
To stabilize the zero-level set of the function $\tvc=\tvc(\state)$, we will  consider a control law of the following form:
\begin{equation}\label{eq:MRfeedback}
    \ac=\ac_\nom(\prj(\state))+\lfb(\prj(\state))\tvc.
\end{equation}
Here $\ac_\nom:\mgspace\to\Ri{m}$ is the known function corresponding to the control curve of the $\mg$-parameterized maneuver (see Def.~\ref{def:s-param_mamever}) and   $\lfb:\mgspace \to \Ri{m\times n}$ is smooth  (i.e. of class $\cont{\infty}$).  Note that, due to $\prj(\cdot)$ being locally Lipschitz in $\prjspace$,  the   (local) existence and uniqueness of a solution $\state(t)$ to \eqref{eq:DynSys}  is guaranteed  if $\ac$ is taken according to  \eqref{eq:MRfeedback}, as the right-hand side of \eqref{eq:DynSys}   is then locally Lipschitz continuous in a neighborhood of $\nomorb$.

Whenever $\jac{\prj}(\cdot)$ is well defined,  we have by the chain rule that the time derivative of $\tvc$ under \eqref{eq:MRfeedback} is given by
\begin{equation}\label{eq:errorDyn}
    \dot{\tvc}=\jac{\tvc}(\state)\left(\ffunc(\state)+\gfunc(\state)\left[\ac_\nom(\prj)+\lfb(\prj)\tvc\right]\right),
\end{equation}
where $\prj=\prj(\state)$. 
With the aim of providing conditions ensuring that a control law of the form \eqref{eq:MRfeedback} is a
solution to Problem~\ref{probl:EOSofPPm}, we state the following lemma, which we later will  use to derive the first-order approximation of the right-hand side of \eqref{eq:errorDyn} with respect to $\tvc$.
\begin{lem}\label{lemma:altFuncRep}
Any $\cont{2}$ function $\sigma:\Ri{n}\to\Ri{}$, satisfying $\sigma(y)=0$  for all $y\in\nomorb$, can be be equivalently rewritten as
\begin{equation}\label{eq:TErewriteFncs}
\sigma(\state)=\jac{\sigma}(\xp(\state))\tvc(\state)+\bigO(\|\tvc(\state)\|^2)
 \end{equation}
 for almost all $\state$
 in a neighborhood $\prjspacesub\subseteq\prjspace$ of  $\nomorb$.
\end{lem}

For  $\Efunc(\cdot)$  as in \eqref{eq:Edef}, let $A_{cl}({{\mg}}):=A_s({{\mg}})+B_s({\mg})\lfb({\mg}) $ with
\begin{equation}\label{eq:AandBdef}
    A_s(\mg):=\Efunc(\state_\nom(\mg),\ac_\nom(\mg)) \ \ \text{and} \ \ B_s(\mg):=\gfunc(\state_\nom(\mg)).
\end{equation}
We may then use Lemma~\ref{lemma:altFuncRep} to state the following.

 \begin{prop}\label{prop:dynamicsOfCoords}
  For some   projection operator $\prj:\prjspace\to\mgspace$ as by Definition~\ref{def:ProjOp},
 consider  the closed-loop system \eqref{eq:DynSys}  under the (locally Lipschitz) control law \eqref{eq:MRfeedback}. There then exists a neighborhood $\neighb(\nomorb)$ of $\nomorb$, such that the time derivative of $\tvc=\tvc(\state)$, defined in \eqref{eq:eTC},  can be written in the following forms within three specific subsets of $\neighb(\nomorb)$: \textit{i})  If $\state(t)\in\mathcal{H}_i\cap  \neighb(\nomorb)$ with  $i\in\{\ini,\fin\}$ fixed,  
 then
 \begin{equation}\label{eq:JacDyn}
     \dot{\tvc}=A_{cl}(\mg_i)\tvc+\bigO(\|\tvc\|^2);
     \end{equation}
     \textit{ii}) If $\state(t) \in\tubespace\cap\neighb(\nomorb)$, then
 \begin{equation}\label{eq:TubeDyn}
    \dot{\tvc}=\left[\tvcjac(\prj)A_{cl}(\prj)-\nomflow(\prj)\projjac'(\prj)\nvel(\prj)\right]\tvcjac(\prj)\tvc+\bigO(\|\tvc\|^2), 
\end{equation}
where $\prj=\prj(\state)$ and $\projjac'(\mg)=\nomflow\transp(\mg)\hess{\prj}(\state_\nom(\mg))$.  
 \end{prop}

Consider  the linear, time-invariant system
\begin{equation}\label{eq:LTI}
    \dot{y}=A_{cl}(\mg_i)y, \quad y\in\Ri{n}, 
\end{equation}
for some fixed $i\in\{\ini,\fin\}$. It 
corresponds to the first-order approximation system of \eqref{eq:JacDyn}. It is also equivalent to the Jacobian linearization of \eqref{eq:DynSys}  under the  linear control law $\ac=\ac_\nom(\mg_i)+\lfb(\mg_i)(\state-\state_i)$
 about the respective equilibrium point.
 The first-order approximation system of \eqref{eq:TubeDyn} along $\nomorb$, on the other hand, is equivalent to the following system of differential-algebraic equations:
\begin{subequations}\label{eq:LTVD}
\begin{align}
        \dot{z}&=\left[\tvcjac(\mg)A_{cl}(\mg)-\nomflow(\mg)\projjac'(\mg)\nvel(\mg)\right]z,
        \\
        0&=\projjac(\mg)z,\label{eq:LTVDtramsvCond}
\end{align}
\end{subequations}
    where $z\in\Ri{n}$,  $\mg=\mg(t)$  solves \eqref{eq:nvel}, and with   the condition $0=\projjac(\mg)z$ obtained directly from  \eqref{eq:TCTaylorExp}  using Lemma~\ref{lemma:tvcjac}  (see also \cite[Sec. 4]{leonov1995local} or \cite[Thm. 7]{saetre2020IFACWC} for alternative derivations of \eqref{eq:LTVD}). Note that \eqref{eq:LTVD} is  different to the first-order variational system of  \eqref{eq:DynSys} about  $\state_\nom(\mg(t))$, which instead is  given by
 \begin{equation}\label{eq:linVarSys}
     \dot{\chi}=\left[A_{cl}(\mg)+B_s(\mg)\left(\ac_\nom'(\mg)-\lfb(\mg)\nomflow(\mg)\right)\projjac(\mg)\right]\chi.
 \end{equation}
 The solutions to \eqref{eq:LTVD} and \eqref{eq:linVarSys} are however related through $z(t)=\tvcjac(\mg(t))\chi(t)$. Hence, by recalling the properties of $\tvcjac(\cdot)$ (see Lem.~\ref{lemma:tvcjac}), it follows that \eqref{eq:LTVD} captures the transverse components of the variational system \eqref{eq:linVarSys}, and is therefore  referred to as a \emph{transverse linearization}.

It is  well known (see, e.g., \cite[Theorem 4.6]{khalil2002nonlinear}) that the origin of \eqref{eq:LTI}  is exponentially stable at both $\mg_{\ini}$ $\mg_{\fin}$ if, and only if, for any $Q_{\ini},Q_{\fin}\in\Mats^n_{\succ 0}$, there exist  $R_{\ini},R_{\fin}\in\Mats^n_{\succ 0}$ satisfying a pair of  algebraic Lyapunov equations (ALEs):
 \begin{subequations}\label{eq:ALEs}
     \begin{align}\label{eq:ALE0}
      A_{cl}\transp(\mg_{\ini})R_{\ini}+R_{\ini} A_{cl}(\mg_{\ini})&=-Q_{\ini}, 
         \\
         A_{cl}\transp(\mg_{\fin})R_{\fin}+R_{\fin} A_{cl}(\mg_{\fin})&=-Q_{\fin}.
     \end{align}
     \end{subequations}
A  similar statement can also be readily obtained for   \eqref{eq:LTVD} by either a slight reformulation of  Theorem~1 in \citep{saetre2019excessiveECC} or from the stronger statements found in \citep{leonov1995local,leonov1990orbital} (see, respectively, Theorem 5.1 and Theorem 1 therein). 

\begin{lem}
    Suppose there exist $\cont{1}$-smooth matrix-valued functions $R,Q_\perp:\mgspace\to\mathbb{M}^n_{\succ 0}$ such that the projected Lyapunov differential  equation (PrjLDE) 
    \begin{align}\label{eq:PLDE}
    \tvcjac\transp\big[&A_{cl}\transp\tvcjac\transp R+R\tvcjac A_{cl}+Q_\perp\big]\tvcjac
    \\ \nonumber
    &+\nvel\tvcjac\transp
    \big[R'-({\projjac'})\transp \nomflow\transp  R-R\nomflow\projjac'\big]
    \tvcjac
    =\0{n}  
    \end{align}
    is satisfied for all $\mg\in\mgspace$ (here the $\mg$-arguments of the functions have been omitted  for brevity).
     Then the time derivative  of the scalar function
         $V_\perp=z\transp R(\mg(t))z$,
     with $z=z(t)$ governed by \eqref{eq:LTVD}, is  $\dot{V}_\perp=-z\transp Q_\perp(\mg(t))z$.  \qed
\end{lem}

Note here that  by \eqref{eq:LTVDtramsvCond} we have  $z\transp R(\mg)z=z\transp R_\perp(\mg)z$ where $R_\perp(\mg):=\tvcjac\transp(\mg)R(\mg) \tvcjac(\mg)$. Due to the fact that $\tvcjac^2(\mg)=\tvcjac(\mg)$, this motivates the following: 
\begin{prop}\label{Prop:Uniqueness}
Let  the $\cont{1}$ function $\nvel:\mgspace\to\Rip$   satisfy $\nvel(\mg_{\ini})=\nvel(\mg_{\fin})\equiv 0$,  $\nvel'(\mg_{\ini})>0$, and $\nvel(\mg)>0$ for all  $\mg\in\interior{(\mgspace)}$.
    Then there exists a  $\cont{1}$ solution  $R:\mgspace\to\Mats_{\succ 0}^{n}$  to \eqref{eq:PLDE} for some smooth $Q_\perp:\mgspace\to\mathbb{M}^n_{\succ 0}$ if, and only if, there exists a unique  $\cont{1}$ solution $R_\perp:\mgspace\to\Mats_{\succeq 0}^{n}$ to
    \begin{align}\label{eq:MLDE}
        \tvcjac\transp(\mg)\big[A_{cl}\transp(\mg)& R_\perp(\mg)        +R_\perp(\mg)A_{cl}(\mg)
        \\ \nonumber
        &+\nvel(\mg)R_\perp'(\mg)
    +Q_\perp(\mg)\Big]\tvcjac(\mg)=\0{n}
    \end{align}
    satisfying $R_\perp(\mg)=\tvcjac\transp(\mg)R_\perp(\mg) \tvcjac(\mg)$  for all $\mg\in\mgspace$.
\end{prop}

\section{Main results}\label{sec:MainSection} 
We now provide  conditions ensuring that a control law of the form \eqref{eq:MRfeedback} is a solution to   Problem~\ref{probl:EOSofPPm}.
\begin{thm}\label{theorem:MR}
 Given a projection operator $\prj(\cdot)$ as by Definition~\ref{def:ProjOp}, consider  the closed-loop system \eqref{eq:DynSys}  under the (locally Lipschitz) control law \eqref{eq:MRfeedback}.
   If there exists a $\cont{1}$-smooth   matrix function  $R:\mgspace\to\Mats_{\succ 0}^n$ such that
 \begin{enumerate}
     \item[\textbf{1.}] for some  $Q_{\ini},Q_{\fin}\in\Mats_{\succ 0}^n$, $R_{\ini}=R(\mg_{\ini})$ and $R_{\fin}=R(\mg_{\fin})$ satisfy the  ALEs \eqref{eq:ALEs};
     \item[\textbf{2.}] for some smooth  $Q_\perp:\mgspace\to\Mats_{\succ 0}^n$,  $R_\perp(\mg):=\tvcjac\transp(\mg)R(\mg) \tvcjac(\mg)$ 
    satisfies \eqref{eq:MLDE} for all $\mg\in\mgspace$;
 \end{enumerate}
 then
 \begin{itemize}
     \item[a)] the final equilibrium,  $\state_{\fin}$,  is 
 asymptotically stable;
     \item[b)]  the one-dimensional manifold $\nomorb$, defined in \eqref{eq:nomOrb}, is  invariant and exponentially stable;
     \item[c)]  there exists a pair of numbers, $\mu,\nu\in\Ri{}_{>0}$, such that the time derivative of the locally Lipschitz  function $V(\state)=\tvc\transp(\state) R(\prj(\state))\tvc(\state) $  satisfies $\dot{V}(\state)\le-\mu V(\state)$ for almost all $\state$
 in     $\mathfrak{T}=\{\state\in\Ri{n}: \ V(\state)<\nu \}$.
 \end{itemize}
\end{thm}

\begin{rem}\label{remark:epsilonProjOp}
    Under a control law \eqref{eq:MRfeedback} satisfying the conditions in Theorem~\ref{theorem:MR}, any solution of \eqref{eq:DynSys}   initialized in vicinity of $\nomorb$   will  converge either directly to the initial equilibrium $\state_{\ini}$, which is rendered partially unstable (a ``saddle''), or onto $\nomorb\backslash\{\state_{\ini}\}$ and then onward  to  $\state_{\fin}$. 
    This implies that the system's states  can get  ``trapped''  if they  enter the region of attraction of $\state_{\ini}$. Indeed,  they will then converge toward $\state_{\ini}$ at an exponential rate, but never enter  into the tube $\tubespace$ from within which they can  converge to $\state_{\fin}$. This issue can be resolved by some ad hoc modification to the controller \eqref{eq:MRfeedback}. For example, one can  limit the codomain of the projection operator used in \eqref{eq:MRfeedback}. For  an operator of the form \eqref{eq:GenProjOp}, this would correspond to ${\prj(\state)}=\argmin_{\mg \in [\mg_{\ini}+\epsilon,\mg_{\fin}]}(\cdot)$ for some sufficiently small $\epsilon>0$. A similar alternative is to let $\epsilon\in[0,\epsilon_M]$ be a bounded dynamic variable, e.g. $\dot{\epsilon}=\lambda_\epsilon \cdot  \text{sign}\big(\delta_\epsilon-\|\state-\state_{\ini}\|\big)$ for small   $\epsilon_M,\delta_\epsilon,\lambda_\epsilon>0$, although the control law will then no longer be truly static in a neighborhood of $\state_{\ini}$. 
\end{rem}

Before we move on to showing how such a feedback can be constructed, we will apply the method to a simple fully-actuated, one-degree-of-freedom system  as to highlight the effect of the projection operator upon the resulting feedback controller.
\begin{exmp}\label{example:DI}
Consider the double integrator 
\begin{equation*}
    \Ddot{q}=u, \quad q(t),u(t)\in\Ri{},
\end{equation*}
with state vector $x=\colvec(q,\dot{q})$.
Starting from rest at $q_{\ini}$, the task is to drive the system to rest at $q_{\fin} \ (>q_{\ini})$ along the curve $\state_\nom(\mg)=\colvec(\mg,\nvel(\mg))$. Here $\mg\in\mgspace:=[q_{\ini},q_{\fin}]$ and $\nvel(\mg):=\kappa(\mg-q_{\ini}) (q_{\fin}-\mg)^2$ for some constant $\kappa>0$. As $\|\nomflow(\mg)\|^2=1+(\nvel'(\mg))^2\ge 1$, this is an $\mg$-parameterization as by Definition~\ref{def:s-param_mamever}.

\begin{figure}
    \centering
    \includegraphics[width=1.\linewidth]{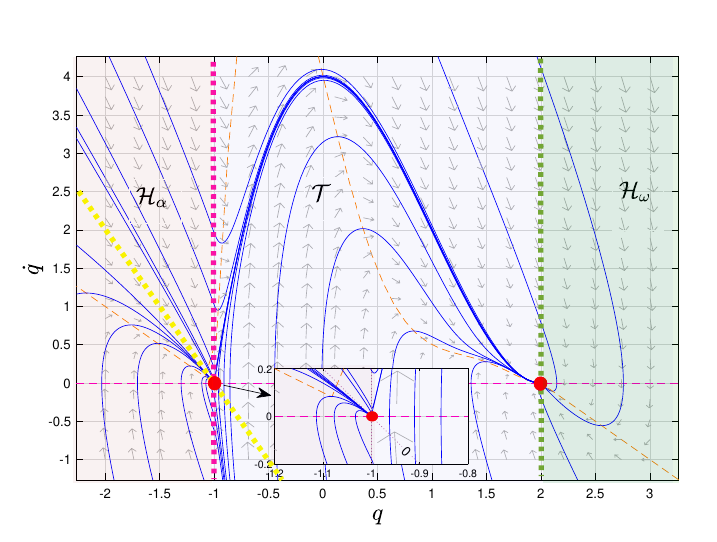}
    \caption{Phase portrait of $\ddot{q}=u$, with $u$ corresponding to Example~\ref{example:DI} for $q_{\ini}=-1$, $q_{\fin}=2$, $\kappa=1$, $p(x)=\text{sat}_{q_{\ini}}^{q_{\fin}}(q)$ and $k_1=k_2=4$. The level curve $\dot{q}+2(q+1)=0$ crossing $(q_{\ini},0)$ is illustrated by the yellow, dotted line. }
    \label{fig:DIexample}
\end{figure}

Suppose $\prj(\cdot)$ is a projection operator in line with  Def.~\ref{def:ProjOp} (we will provide some candidates for this operator shortly).
Using $\prj=\prj(\state)$, we  define $\tvc_1:=q-\prj$, $\tvc_2:=\dot{q}-\nvel(\prj)$ and $\ac_\nom(\prj):=\nvel'(\prj)\nvel(\prj)$, such that $u=\ac_\nom(\prj)-k_1\tvc_1-k_{2}\tvc_2$ is of the form of \eqref{eq:MRfeedback}.
 Let us therefore check when this feedback, corresponding to a constant $K=[-k_1,-k_2]$, satisfies the conditions in Theorem~\ref{theorem:MR} for a given $\prj(\cdot)$.  
 
Let $k_1,k_2>0$  such that 
     $A_{cl}:={\tiny\begin{bmatrix}0 & 1 \\ -k_1 & -k_2 \end{bmatrix}}$
      is Hurwitz, and denote by $R\in\Mats_{\succ 0}^{2}$   the  unique  solution to $A_{cl}\transp R+RA_{cl}=-2\I{2}$, which corresponds to the ALEs \eqref{eq:ALEs}. We may then consider the (locally Lipschitz) Lyapunov function  candidate $V=2\inv\tvc\transp R\tvc$, with $\tvc=\colvec(\tvc_1,\tvc_2)$, whose zero-level set evidently corresponds to the desired orbit. Within the interiors of $\Pi(q_{\ini})$ and $\Pi(q_{\fin})$, with $\Pi(\cdot)$ defined in \eqref{eq:MPS},  we  have $\dot{V}=-\|\tvc\|^2$ since  $\|\jac{\prj}\|=0$ therein. To determine the stability of the orbit as a whole,  we therefore need to check that we also have contraction within some tubular neighborhood contained in $\tubespace$ for the chosen  projection operator.  We  consider two different such operators next. 

 By taking inspiration from   \citep{hauser1995maneuver}, let us first consider the projection operator  corresponding to taking $\Lambda=R$ in \eqref{eq:GenProjOp}. Using \eqref{eq:genProjOpJac}, we then observe  that $\tvcjac\transp(\mg) R\nomflow(\mg)=\0{1\times 2}$ for all $\mg\in\mgspace$. Hence \eqref{eq:MLDE} is everywhere satisfied for this $R_\perp$ and $Q=\I{2}$ (see  \cite{hauser1995maneuver} for further details). Moreover, we have $\dot{V}=-\|\tvc\|^2$ for all $\state$ such that $\prj(\state)\in(q_{\ini},q_{\fin})$.

Consider now instead the operator obtained by taking $\Lambda=\text{diag}(1,0)$ in \eqref{eq:GenProjOp}. This is equivalent to  $\prj(x)=\sat_{q_{\ini}}^{q_{\fin}}(q)$, where $\sat_a^b(q)=\max(a,\min(q,b))$ is the saturation function  (an  example of using this operator is shown in Figure~\ref{fig:DIexample}).
Clearly then $\tvc_1\equiv 0$ for $q\in[q_{\ini}, q_{\fin}]$, while it can be  shown that  $\dot{\tvc}_2=-(k_2+\nvel'(\prj))\tvc_2$.
Thus,   the time derivative of the above Lyapunov function candidate  satisfies
\begin{equation*}
  \dot{V}=-R_{22}(k_2+\nvel'(\prj))\tvc_2^2=-R_{22}(k_2+\nvel'(\prj))\|\tvc\|^2  
\end{equation*}
 whenever $q\in(q_{\ini},q_{\fin})$, with $R_{22}>0$ the bottom-right element of $R$. We may therefore ensure that $V$ will be strictly decreasing everywhere inside the tube (except, of course, on the nominal orbit) by taking, e.g., $k_2>\sup_{\mg\in \mgspace}|\nvel'(\mg)|$.  This is nevertheless in contrast to the previous operator (i.e. $\Lambda=R$) where $k_2>0$ could be taken arbitrarily small and still ensure contraction, thus highlighting the dependence of the feedback $K(\cdot)$ upon the choice of $\prj(\cdot)$.
\end{exmp}

As shown in this example, it is trivial to combine Theorem~\ref{theorem:MR} with a specific projection operator as in \cite{hauser1995maneuver}:
\begin{cor}\label{cor:Hauser}
If there exist $\cont{1}$-smooth matrix-valued functions $R,Q:\mgspace\to\Mats_{\succ 0}^n$ satisfying, for all $\mg\in\mgspace$,
\begin{equation}
    \nvel(\mg)R'(\mg)+A_{cl}\transp(\mg)R(\mg)+R(\mg) A_{cl}(\mg)=-Q(\mg),
\end{equation}
then $R(\mg)$ satisfies the conditions in Theorem~\ref{theorem:MR} provided that $\state_\nom(\cdot)$ is $\cont{3}$ and $\prj(\cdot)$ is taken as in Proposition~\ref{prop:GenProjOp} with $\Lambda(\mg)=R(\mg)$.  \qed
\end{cor}

Corollary~\ref{cor:Hauser}  shows the possibility of finding a feedback matrix $\lfb(\cdot)$ that solves Problem~\ref{probl:EOSofPPm} by solving a differential Riccati equation. However, it also forces one to use a particular projection operator (see Prop.~\ref{prop:GenProjOp}), which generally requires one to solve an optimization problem at each iteration. Meanwhile, Example~\ref{example:DI} showed that it also can be possible to find projection operators which are very simple and can be computed directly. This motivates  a method which allows one to attempt to find  a solution for any choice of projection operator. To this end, let $B_\perp(\mg):=\tvcjac(\mg)B_\mg(\mg)$ and
\begin{align*}
    A_\perp(\mg):=\tvcjac(\mg) A_\mg(\mg) -\nvel(\mg) \nomflow(\mg)\nomflow\transp(\mg)\hess{\prj}(\state_\nom(\mg))\tvcjac(\mg)   .
\end{align*}
Inspired by   linear matrix inequality (LMI) approaches such as that  in \citep{bernussou1989linear},  the following statement provides one such method.

\begin{prop}\label{prop:LMIcontDesign}
Given a projection operator $\prj(\cdot)$ in the sense of Definition~\ref{def:ProjOp},
  suppose that for a strictly positive,  smooth function  $\lambda:\mgspace\to\Ri{}_{>0}$, there exists a pair of smooth matrix-valued functions
  $Y:\mgspace \to \Ri{m\times n}$ and $W:\mgspace\to\mathbb{M}^n_{\succ 0}$,
  which for all $\mg\in\mgspace$ satisfy the matrix inequality
\begin{align}\label{eq:DLMI}
    &\nvel(\mg) W'(\mg) -
     W(\mg)A_\perp\transp(\mg)- A_\perp(\mg) W(\mg)
    -Y\transp(\mg) B_\perp\transp(\mg)
     \nonumber  \\
    &- B_\perp(\mg) Y(\mg) 
    -\lambda(\mg)[\tvcjac (\mg)W(\mg)+W(\mg)\tvcjac\transp(\mg)]\succeq \0{n}.
\end{align}
Further suppose that for some $K_{\ini},K_{\fin}\in\Ri{m\times n}$ which are such that $(A_\mg(\mg_{\ini})+B_\mg(\mg_{\ini})K_{\ini})$ and $(A_\mg(\mg_{\fin})+B_\mg(\mg_{\fin})K_{\fin})$ are both Hurwitz, the following two identities hold:
\begin{equation}\label{eq:EqContConstraints}
    K_{\ini}W(\mg_{\ini})=Y(\mg_{\ini}) \quad \text{and} \quad K_{\fin}W(\mg_{\fin})=Y(\mg_{\fin}).
\end{equation}
 Then by taking $\lfb(\mg)=Y(\mg)W\inv(\mg)$ in \eqref{eq:MRfeedback}  the matrix function $R(\mg)=W\inv(\mg)$ satisfies all the requirements stated in Theorem~\ref{theorem:MR}.
\end{prop}

In order to find a solution  pair $(W,Y)$  to Proposition~\ref{prop:LMIcontDesign}, one can use some transcription method as to discretize the differential LMI \eqref{eq:DLMI} into a finite set of LMIs. One can then attempt to find an approximate solution using semidefinite programming (SDP). In regard to handling the constant stabilizing matrices $K_{\ini}$ and $K_{\fin}$ in the resulting  SDP  formulation, there are two main options: \newline
    \textbf{1)} Add, for both  $\mg\in\{\mg_{\ini},\mg_{\fin}\}$, the LMI constraints
    \begin{equation*}
     W(\mg)A_\mg\transp(\mg)+A_\mg(\mg)W(\mg)+Y\transp(\mg)B_\mg(\mg)+B_\mg(\mg)Y(\mg)\prec \0{n} ;
    \end{equation*}
    \textbf{2)}
    Add the equality constraints \eqref{eq:EqContConstraints}, in which some stabilizing matrices $K_{\ini}$ and $K_{\fin}$ have already been found.

    In case of the latter option, one can for example use LQR: Take, for both $i\in\{\ini,\fin\}$, $K_i=-\Gamma_i\inv B_\mg\transp(\mg_i)R_i$, where $R_i\in\Mats^{n}_{\succ 0}$ solves  the  algebraic Riccati equation
\begin{equation}\label{eq:AREs}
    A_\mg\transp(\mg_i)R_i+R_iA_\mg(\mg_i)-R_iB_\mg(\mg_i)\Gamma_i\inv B_\mg\transp(\mg_i)R_i=-Q_i
\end{equation}
  given some  $\Gamma_i\in\Mats^{m}_{\succ 0}$ and $Q_i\in\Mats^{n}_{\succ 0}$.
\section{Planning point-to-point maneuvers of underactuated  mechanical systems}\label{sec:UnAcSys}
Consider now the following task: Find an  $\mg$-parameterized  PtP maneuver (see Def.~\ref{def:s-param_mamever}) of an underactuated mechanical systems with $n_q$ degrees of freedom, one degree of underactuation, and equations of  motion
\begin{equation}\label{eq:Manipulator equation}
	\mathbf{M}(\gc)\ddot{\gc}+\mathbf{C}(\gc,\dot{\gc})\dot{\gc}+\mathbf{G}(\gc)=\mathbf{B}_u\ac.
\end{equation}
Here  $\gc=\colvec(\gc_1,\dots,\gc_{n_q})\in\Ri{n_q}$ are  generalized coordinates, $\dot{\gc}\in\Ri{n_q}$  the corresponding generalized velocities, $\state=\colvec(\gc,\dot{\gc})$ denotes the $n=2n_q$ states, while $\ac \in \Ri{m}$ is a vector of $m=n_q-1$  control inputs; $\mathbf{M}(\cdot)\in\Mats_{\succ 0}^{n_q}$ is the (smooth)    inertia matrix;  the constant matrix  $\mathbf{B}_u\in\Ri{n_q\times m}$ has full rank;
 $\mathbf{C}(\cdot,\cdot)$ corresponds  to Coriolis and centrifugal forces, which we in this paper write as $\mathbf{C}(\gc,\dot{\gc})=\mathbf{C}_1(\gc,\dot{\gc})+\mathbf{C}_2(\gc,\dot{\gc})$ with
 $\mathbf{C}_1(\gc,\dot{\gc}):=\sum_{i=1}^{n_q}\frac{\partial \mathbf{M}(\gc)}{\partial q_i}\dot{q}_i $ and
 $\mathbf{C}_2(\gc,\dot{\gc}):=-\frac{1}{2}\begin{bmatrix} \frac{\partial \mathbf{M}(\gc)}{\partial q_1}\dot{\gc},\dots, \frac{\partial \mathbf{M}(\gc)}{\partial q_{n_q}} \dot{\gc} \end{bmatrix}\transp$;
while $\mathbf{G}(\cdot)\in\Ri{n_q}$ is the (smooth) gradient of the system's potential energy. 

For a pair of points (configurations) $\gc_{\ini}$ and $\gc_{\fin}$, $\gc_{\ini}\neq \gc_{\fin}$,  suppose there exist $\ac_{\ini},\ac_{\fin}\in\Ri{m}$ such that $\mathbf{G}(\gc_{\ini})\equiv \mathbf{B}_u \ac_{\ini}$ and $\mathbf{G}(\gc_{\fin})\equiv \mathbf{B}_u\ac_{\fin}$.
The task we want solve in this section can then be more accurately  formulated:
\begin{prob}\label{problem:ManeuverGeneration}
 For $\state_{\ini}=\colvec(\gc_{\ini},\0{n_q\times 1})$ and $\state_{\fin}=\colvec(\gc_{\fin},\0{n_q\times 1})$, find  for the system \eqref{eq:Manipulator equation} an $\mg$-parameterized PtP maneuver connecting $\state_{\ini}$ and $\state_{\fin}$, i.e.,
a triplet $(\state_\nom,\ac_\nom,\nvel)$ of the form \eqref{eq:maneuverTriplet} satisfying Definition~\ref{def:s-param_mamever}.
\end{prob}

To solve this problem, we propose  a procedure inspired by the approach  in \cite{shiriaev2005constructive}.

\subsection{Synchronization function--based orbit generation}
Since \eqref{eq:Manipulator equation} is a second-order system, the state curve $\state_\nom:\mgspace\to\nomorb$    (see Def.~\ref{def:s-param_mamever}) can be written on the form
\begin{equation}\label{eq:MGnominalTrajectory}
     \state_\nom(\mg):=\colvec\big(\vc(\mg),\vc'(\mg)\nvel(\mg)\big).
\end{equation}
Here   $\vc(\mg)=\colvec(\vcs_1(\mg),\dots,\vcs_{n_q}(\mg))$ is a  vector-valued function, which we will assume is smooth, that traces out a curve in the configuration space of the system. As one may consider the generalized coordinates as being synchronized when confined to this curve, we will refer to the smooth, scalar functions $\vcs_i(\cdot)$ as \emph{synchronization functions}.\footnote{If one replaces $\mg$ with a known function of only the generalized coordinates, i.e. $\theta=\theta(\gc)$, then the relations $\vcs_i(\theta)$ have commonly been referred to as \emph{virtual (holonomic) constraints} (see, e.g., \cite{shiriaev2005constructive}). This terminology is somewhat misleading for the purpose we consider in this paper, however, and we therefore use the more fitting notion of synchronization functions.}  Moreover, the scalar function $\nvel:\mgspace\to\Rip $ may now, in addition to governing the dynamics of the curve parameter $\mg$ (see \eqref{eq:nvel}), also be considered as to set the speed at which the curve formed by $\vc(\cdot)$ is traversed. 

Let us  now derive condition upon the functions  $\vc(\cdot)$, $\nvel(\cdot)$
 and $\ac_\nom(\mg)$ such that they together provide as solution to Problem~\ref{problem:ManeuverGeneration}. In this regard, we first note that  Property  \textbf{P4} in Definition~\ref{def:s-param_mamever}, i.e.  $\|\nomflow(\mg)\|>0$, is equivalent to
\begin{align}\label{eq:nonVanCondUndAcSys}
\|\vc'(\mg)\|^2+\|\vc''(\mg)\nvel(\mg)+\vc'(\mg)\nvel'(\mg)\|^2>0.
\end{align}
Next we note that Property \textbf{P2} obviously requires that  $\vc(\mg_{\ini})=\gc_{\ini}$ and $\vc(\mg_{\fin})=\gc_{\fin}$.
Furthermore, to ensure consistency with the dynamics of \eqref{eq:Manipulator equation}, corresponding to Property \textbf{P5},  it is clear that the functions  $\vc(\cdot)$, $\nvel(\cdot)$
 and $\ac_\nom(\mg)$ must satisfy the following equality for all $\mg\in\mgspace$:
\begin{equation}   \label{eq:constrainedDynamics}
    \balpha(\mg)\nvel'(\mg)\nvel(\mg)+\bbeta(\mg)\nvel^2(\mg)+\bgamma(\mg)=\mathbf{B}_u\ac_\nom(\mg).
\end{equation}
Here  $\balpha(\mg):=    \mathbf{M}\big(\vc(\mg)\big)\vc'(\mg)$,  $\bbeta(\mg):=\mathbf{M}\big(\vc(\mg)\big)\vc''(\mg)+\mathbf{C}\big(\vc(\mg),\vc'(\mg)\big)\vc'(\mg)$, and $\bgamma(\mg):=\mathbf{G}\big(\vc(\mg)\big)$.
Due to the assumption that $\mathbf{B}_\ac$ has full rank, we can multiply \eqref{eq:constrainedDynamics} from the left by any of its  left inverses $\mathbf{B}_\ac\pinv\in\Ri{m\times n_q}$ i.e. $\mathbf{B}_\ac\pinv\mathbf{B}_\ac=\I{m}$, to obtain
\begin{equation}\label{eq:nomCont}
   \ac_\nom(\mg)=\mathbf{B}_u\pinv\left[\balpha(\mg)\nvel'(\mg)\nvel(\mg)+\bbeta(\mg)\nvel^2(\mg)+\bgamma(\mg)\right].
\end{equation}
Hence, if $\vc:\mgspace\space\to\Ri{n_q}$ and $\nvel:\mgspace\to\Rip$ are known, then the corresponding  $\ac_\nom(\cdot)$ can be found from \eqref{eq:nomCont}. 

From the above it is clear  that  if the system \eqref{eq:Manipulator equation} was fully actuated, i.e. $m\equiv n_q$, and therefore $\mathbf{B}_u\pinv=\mathbf{B}_u\inv$, then  Property \textbf{P5} would immediately be satisfied simply by taking $\ac_\nom(\cdot)$ according to \eqref{eq:nomCont} for
any combination of $\vc(\cdot)$ and $\nvel(\cdot)$  (see  Example~\ref{example:DI}). 
This is, however, not the case for the underactuated systems we consider, as  $\mathbf{B}_u\in\Ri{n_q\times n_q-1}$   has a family of  full-rank left annihilators. Denote by  $\mathbf{B}_u^\perp\in\Ri{1\times n_q}$  such an annihilator, i.e. $\mathbf{B}_u^\perp\mathbf{B}_u=\0{1\times m}$.  Multiplying  \eqref{eq:constrainedDynamics} from the left by $\mathbf{B}_u^\perp$, one then finds that $\vc(\cdot)$ and $\nvel(\cdot)$ must satisfy 
\begin{equation}
\label{eq:ABG}
    \alpha(\mg)\nvel'(\mg)\nvel(\mg)+\beta(\mg)\nvel^2(\mg)+\gamma(\mg)=0
\end{equation}
for all $\mg\in\mgspace$, where  $\alpha(\mg):=\mathbf{B}_u^{\perp}\balpha(\mg)$, $\beta(\mg):=\mathbf{B}_u^{\perp}\bbeta(\mg)$ and $\gamma(\mg):=\mathbf{B}_u^{\perp}\bgamma(\mg)$.

Our suggested approach for solving Problem~\ref{problem:ManeuverGeneration} can now roughly be described as follows: For a particular choice of a smooth $\vc(\cdot)$, try to find some $\nvel(\cdot)$ satisfying \eqref{eq:ABG} and Property \textbf{P3} in Definition~\ref{def:s-param_mamever}, i.e. $\nvel(\mg_{\ini})=\nvel(\mg_{\fin})\equiv 0$ and $\nvel(\mg)>0$  for all $\mg\in\interior{(\mgspace)}$. If a  (satisfactory) solution $\nvel(\cdot)$ is found, then the corresponding unique $\ac_\nom(\cdot)$ is in turn found directly from \eqref{eq:nomCont}.

In order to help us find such a function $\nvel(\cdot)$, we will utilize the fact that  a solution $\mg=\mg(t)$ to $\dot{\mg}=\nvel(\mg)$  must then also be a solution
to the second-order differential equation (cf. \eqref{eq:ABG})
\begin{equation}\label{eq:reducedDynamics}
     \alpha(\mg)\ddot{s}+\beta(\mg)\dot{s}^2+\gamma(\mg)=0.
 \end{equation}
 We will refer to \eqref{eq:reducedDynamics} as the \emph{reduced dynamics} associated with the synchronization functions $\vc(\cdot)$. Next we briefly review some  key properties of this equation, originally derived in \cite{shiriaev2005constructive,shiriaev2006periodic}.

\subsection{Properties of the reduced dynamics}
The following is a  (weaker)  reformulation of Theorem 3 in \cite{shiriaev2006periodic}, and thus stated without proof.
\begin{lem}\label{lemma:abg-eqs}
 Let $\mg_e\in\mgspace$ be an equilibrium point  of \eqref{eq:reducedDynamics},
 i.e. $\gamma(\mg_e)\equiv 0$, satisfying   $\alpha(\mg_e)\neq0$, and denote
\begin{equation}\label{eq:EqnuFunc}
    \nu(s):={\gamma'(\mg)}/{\alpha(\mg)}.
\end{equation}
 Then the equilibrium point $\mg_e$ is a center  if $\nu(\mg_e)>0$, while it is a saddle if $\nu(\mg_e)<0$. \qed
\end{lem}
Here the conditions for a saddle equilibrium follows directly from the Hartman--Grobman theorem (see also \cite[Sec. 20]{hahn1967stability}), whereas the condition for a center equilibrium point, on the other hand, can be attained by noticing that the solutions of \eqref{eq:ABG} form certain level curves. More precisely, let $\nvel(\cdot)\ge 0$ solve \eqref{eq:ABG}, and note that 
$\beta(\mg):=\alpha'(\mg)+\hat{\beta}(\mg)$ with $\hat{\beta}(\mg):=\mathbf{B}_u^{\perp}\mathbf{C}_2(\vc(\mg),\vc'(\mg))\vc'(\mg)$. Then
\begin{equation}\label{eq:PsiDef}
    \frac{1}{2}\alpha(\mg)\exp\left({\int_{\mg_r}^{\mg}\frac{2\hat{\beta}(\eta)}{\alpha(\eta)}d\eta}\right)  =:\frac{1}{2}\alpha(\mg) \Psi(\mg_r,\mg)    
\end{equation}
is an integrating factor  of \eqref{eq:ABG} for any $\mg_r\in\mgspace$.
By  \cite[Thm. 1]{shiriaev2005constructive}, if $\mg=\mg(t)\in\mgspace$ is simultaneously a solution to \eqref{eq:reducedDynamics} and to $\dot{\mg}=\nvel(\mg)$, with  $\nvel:\mgspace\to\Rip$ strictly positive on $\interior{(\mgspace)}$, then  for any pair of points $\mg_1,\mg_2\in\mgspace$:
\begin{align}\label{eq:IntOfABG}
    {\alpha^2(\mg_2)}\nvel^2(\mg_2)-&\Psi(\mg_2,\mg_1)\Big[{\alpha^2 (\mg_1)}\nvel^2(\mg_1)
    \\ \nonumber
    &-2\int_{\mg_1}^{\mg_2}\Psi(\mg_1,\tau)\alpha(\tau)\gamma(\tau)d\tau\Big]=0.
\end{align}
Note that for certain systems, $\hat{\beta}(\mg)\equiv 0$ $\forall\mg\in\mgspace$, and hence $\Psi\equiv 1$.
 This property, which  can make it significantly easier to check if \eqref{eq:IntOfABG} is satisfied, holds for all systems whose inertia matrix $\mathbf{M}(\cdot)$ is constant, and for any system where the passive  joint is the first in a kinematic chain, such as underactuated systems of Class-I according to the classification of \cite{olfati2001nonlinear}.

\subsection{Conditions for the existence of a PtP maneuver}\label{sec:condPtPmanUMS}
We will now demonstrate how one can use the properties of the reduced dynamics in order to obtain a solution to Problem~\ref{problem:ManeuverGeneration}.
In this regard, recall the definitions of $\nu(\cdot)$ and $\Psi(\cdot)$ given in  \eqref{eq:EqnuFunc} and \eqref{eq:PsiDef}, respectively. 
\begin{thm}\label{theorem:PtPcondsUAS}
 Let the smooth vector-valued function $\vc:\mgspace\to\Ri{n_q}$ be such that
 $\vc(\mg_{\ini})=\gc_{\ini}$, $\vc(\mg_{\fin})=\gc_{\fin}$, $ \|\vc'(\mg_{\ini})\|\neq 0$,
$\|\vc'(\mg_{\fin})\|\neq 0$, $\nu(\mg_{\ini})\le 0$ and $\nu(\mg_{\fin})\le 0$.
 Further suppose that the following conditions hold: $\alpha(\mg)\neq 0$ for all   $\mg\in\mgspace$; there exists a single point $\mg_e\in\interior{(\mgspace)}$   satisfying $\gamma(\mg_e)\equiv0$, for which $\nu(\mg_e)>0$; 
  and  
    \begin{equation}\label{eq:C1cond}
        \int_{\mg_{\ini}}^{\mg_{\fin}}\Psi(\mg_{\ini},\tau)\alpha(\tau)\gamma(\tau)d\tau\equiv 0.
    \end{equation}
    Then there exists a unique, bounded, smooth function $\nvel:\mgspace\to\Rip$ satisfying $\eqref{eq:ABG}$, such that the triplet $(\state_\nom,\ac_\nom,\nvel)$, with $\state_\nom(\cdot)$ given by \eqref{eq:MGnominalTrajectory} and $\ac_\nom(\cdot)$ by \eqref{eq:nomCont}, is a solution to Problem~\ref{problem:ManeuverGeneration}. That is, they constitute an $\mg$-parameterized point-to-point maneuver of  \eqref{eq:Manipulator equation} as by Definition~\ref{def:s-param_mamever}. 
\end{thm}
\begin{rem}\label{rem:PtPmanHyperbolic}
As $\|\mathbf{G}(\gc_{\ini})-\mathbf{B}_u\ac_{\ini}\|=\|\mathbf{G}(\gc_{\fin})-\mathbf{B}_u\ac_{\fin}\|=0$, a solution to Theorem~\ref{theorem:PtPcondsUAS}   implies  $\gamma(\hat{\mg})\equiv 0$ for $\hat{\mg}\in\{\mg_{\ini},\mg_{\fin}\}$. Hence \eqref{eq:ABG}  is then trivially true at $\hat{\mg}\in\{\mg_{\ini},\mg_{\fin}\}$, while from its derivative  with respect to $\mg$,
\begin{align*}
    \alpha \nvel''\nvel+\alpha(\nvel')^2+\big(3\alpha'+2\hat{\beta}\big)\nvel'\nvel
    +\big(\alpha''+\hat{\beta}'\big)\nvel^2+\gamma'=0,
\end{align*}
one finds that $(\nvel'(\hat{\mg}))^2=-\gamma'(\hat{\mg})/\alpha(\hat{\mg})$. Thus, for $\mg_{\ini}$ and $\mg_{\fin}$ to be  hyperbolic (saddle) equilibrium points of \eqref{eq:reducedDynamics},  and consequently   $\nvel'(\mg_{\ini})>0$ and $\nvel'(\mg_{\fin})<0$, it is  further required that $\nu(\mg_{\ini})<0$ and $\nu(\mg_{\fin})<0$.
 From this, one can deduce that the function $\gamma(\mg)/\alpha(\mg)$ then  must change its sign an odd number of times over the open interval $(\mg_{\ini},\mg_{\fin})$. Considering only one sign change, the necessary existence of   a point $\mg_e\in\interior{(\mgspace)}$ for which $\gamma(\mg_e)=0$ and $\nu(\mg_e)>0$ (i.e. a center) is evident.
\end{rem}
\begin{rem}\label{rem:SurovSingPointCond}
Due to the requirement of a center on $\interior{(\mgspace)}$, 
Theorem~\ref{theorem:PtPcondsUAS}  cannot be used to   construct an $\mg$-parameterized PtP maneuver between two adjacent equilibria for 
systems where the  equilibria of \eqref{eq:reducedDynamics}  are fixed. In light of Remark~\ref{rem:PtPmanHyperbolic}, one can in such cases instead attempt to use an alternative set of conditions which are based on $\alpha(\mg)$ changing its sign once over $\interior{(\mgspace)}$ instead of $\gamma(\mg)$. Such conditions can be   obtained from  Theorem~1 in  \cite{surov2018new}, and correspond to  replacing the conditions in the second sentence  in Theorem~\ref{theorem:PtPcondsUAS} with the following: 
 $\nu(\mg_{\ini})< 0$ and $\nu(\mg_{\fin})<0$;   $\gamma(\mg)> 0$ for all $\mg\in\interior{(\mgspace)}$;  and there exists a single point $\mg_s\in\interior{\mgspace}$ satisfying $\alpha(\mg_s)\equiv 0 $ and $\hat{\beta}(\mg_s)<-\frac{3}{2}\alpha'(\mg_s)<0$. Roughly speaking, these conditions ensure that the point $(\mg,\dot{\mg})=\big(\mg_s,\sqrt{-\gamma(\mg_s)/\beta(\mg_s)}\big)$ is finite-time attractive (resp. repellent) for all solutions of \eqref{eq:reducedDynamics} within a neighborhood lying to the left (resp. right) of this point in the upper $(\mg,\dot{\mg})$-plane.
\end{rem}

\section{ Application to non-prehensile manipulation}\label{sec:NPexampleSection}
We will now apply both the motion planning method proposed in Section~\ref{sec:UnAcSys} and the feedback design approach outlined in Section~\ref{sec:MainSection} as  to solve the following non-prehensile manipulation  \citep{ruggiero2018nonprehensile}  problem:  Generate an asymptotically orbitally stable  PtP  motion  corresponding to a ball rolling between any two points upon an actuated planar frame. We begin by describing the system model and provide some necessary assumptions.

\subsection{System description and mathematical model}\label{sec:SysDesvNonPre}
\begin{figure}
    \centering
    \includegraphics[width=0.7\linewidth]{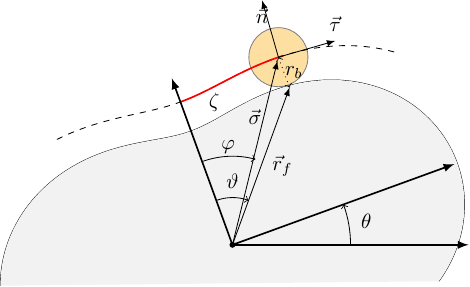}
\caption{The coordinate convention used in Section~\ref{sec:SysDesvNonPre}, with frame having the form of the ``butterfly'' robot.}
    \label{fig:BRmain}
\end{figure}
Consider a ball of (effective) radius $r_b$ which is rolling without slipping upon the boundary of an actuated frame; see  Figure~\ref{fig:BRmain}. The edge of the frame is traced out by the polar coordinates $(\vartheta,r_f(\vartheta))$, with $\vartheta\in\mathcal{I}\subseteq\mathbb{S}^1$ and where the scalar function $r_f:\mathcal{I}\to\Ri{}_{>0}$ is smooth. 
This representation can be used to describe several well-known nonlinear systems, including the {ball-and-beam}  \citep{hauser1992nonlinear}, $r_f(\vartheta)=\frac{\text{const.}}{\cos(\vartheta)}$;  the {disk-on-disk} \citep{ryu2013control}, $r_f(\vartheta)=\text{const.}$;
 as well as  the so-called  {``butterfly'' robot} \citep{lynch1998roles},  whose frame, as in \cite{surov2015case}, can   be of the form
 \begin{equation}\label{eq:BRframe}
     r_f(\vartheta)=a-b\cos(2\vartheta), \quad a,b\in\Ri{}_{>0}.
 \end{equation}
We will make the following assumptions, whose validity  must be checked for any found motion of the system:
\begin{itemize}
    \item[\bf A1.] The ball's center traces out a smooth curve when it traverses the frame;\footnote{Mathematically, this is equivalent to $r_b\kappa_f(\vartheta)<1$   $\forall \vartheta\in\mathcal{I}$, where   $\kappa_f(\vartheta)$ is the signed curvature of the planar curve at $\vartheta$.} 

    \item[\bf A2.]  The ball is always in contact with the frame; 
    \item[\bf A3.]  The ball always rolls without slipping.
\end{itemize}

Let $\theta$ and $\varphi$ be  defined as shown in Figure~\ref{fig:BRmain}, and take $\gc=\colvec(\theta,\varphi)$. Then, in light of the above assumptions, the  system matrices corresponding to \eqref{eq:Manipulator equation} are given by
  \begin{align*}
    \mathbf{M}(\gc)&=\begin{bmatrix} J_f+J_b+m\|\vec{\sigma}\|^2 & -\big(m\vec{\sigma}\cdot\vec{n}+\frac{J_b}{r_b}\big)\zeta'\\
    -\big(m\vec{\sigma}\cdot\vec{n}+\frac{J_b}{r_b}\big)\zeta' & \big(\frac{J_b}{r_b^2}+m\big){\zeta'}^2 \end{bmatrix}  ,
     \\
    \mathbf{C}(\gc,\dot{\gc})&=\begin{bmatrix}
        c_{11}\dot{\varphi} & c_{11}\dot{\theta}-c_{12}\dot{\varphi}\\
        -c_{11}\dot{\theta} & \Big(\frac{J_b}{r_b^2}+m\Big)\zeta'\zeta''\dot{\varphi}
        \end{bmatrix},  \quad \mathbf{B}_u=\begin{bmatrix} 1 \\ 0 \end{bmatrix},
        \\
         \mathbf{G}(\gc)&=\colvec\Big(m\vec{g}\cdot\left(\right(\frac{d}{d\theta}\text{Rot}(\theta)\left)\vec{\sigma}\right),
    m\vec{g}\cdot(\text{Rot}(\theta)\vec{\tau}\zeta')\Big)
  \end{align*}
where $c_{11}:=m\zeta'\vec{\sigma}\cdot\vec{\tau}$, $c_{12}:=\big( m\vec{\sigma}\cdot\vec{n}+\frac{J_b}{r_b}\big)\zeta''+c_{11}\kappa {\zeta'}$ and $\vec{g}=\colvec(0,g)$. See \cite{surov2015case} for a more detailed description of the system parameters and variables, albeit with a slightly different notation.

\subsection{Maneuver design}
We will now utilize the procedure outlined in Section~\ref{sec:UnAcSys} to  plan PtP maneuvers for such systems.
For this purpose, let $\psi(\varphi)$ denote the \emph{tangential angle} of the polar curve at $\varphi$. Namely, the angle such  that the unit tangent vector $\vec{\tau}$ at $\varphi$ can be written as $\vec{\tau}=\colvec(\cos(\psi),\sin(\psi))$; or equivalently, the angle such that $\frac{\partial \psi}{\partial \zeta}=\kappa$ where $\zeta$ is the arc length  and  $\kappa=\kappa(\varphi)$ is the signed curvature of the curve traced out by the ball. Hence $\psi$ is  trivial for systems with constant curvature, e.g., $\psi\equiv 0$ for the ball-and-beam system and $\psi=-\varphi$ for the disk-on-disk. 

With this in mind, consider 
\begin{equation}\label{eq:BRvc}
    \vc(\mg)=\colvec\big(\Theta(\mg)-\psi(\mg),\mg \big), \quad \mg\in\mgspace\subseteq \mathbb{S},
\end{equation}
for some smooth, scalar function $\Theta(\cdot)$.
Simply put, if one takes $\Theta=0$, then the synchronization function \eqref{eq:BRvc}  aligns $\vec{\tau}$ with the fixed horizontal axis (see Figure~\ref{fig:BRmain}), such that the ball can be consider as to be rolling on a horizontal  surface. The function $\Theta(\cdot)$ can therefore be used to slow down or speed up the rolling motion by altering the  ``slope'' upon which the ball rolls.

For this choice of $\vc(\cdot)$, the functions $\alpha(\cdot)$ and $\gamma(\cdot)$ in \eqref{eq:ABG} are given by  $\gamma(\mg)=m g\zeta'\sin({\Theta}(\mg)) $ and
\begin{align*}
    \alpha(\mg)=&\left(\frac{J_b}{R}\left(\kappa+\frac{1}{R}\right)+m(1+\vec{\sigma}\cdot\vec{\kappa})\right){\zeta'}^2 \\
    &-\left(m\vec{\sigma}\cdot\vec{n}+\frac{J_b}{R}\right)\zeta'{\Theta}'.
\end{align*}
From this and Lemma~\ref{lemma:abg-eqs}, the following can be deduced:
\begin{prop}\label{prop:BRprop}
A point $\mg_e\in\mgspace$, for which $\alpha(\mg_e)\neq 0$, is an equilibrium point of \eqref{eq:reducedDynamics} if $\Theta(\mg_e)\equiv 0$. Moreover, it is a center if ${\Theta}'(\mg_e)/\alpha(\mg_e)>0$, or a saddle if ${\Theta}'(\mg_e)/\alpha(\mg_e)<0$. \qed
\end{prop}

One can therefore choose the equilibrium points of \eqref{eq:reducedDynamics} freely through the choice of $\Theta$. 
In light of the discussion in Section~\ref{sec:condPtPmanUMS}, this in turn can be utilized to find a solution satisfying the conditions in  Theorem~\ref{theorem:PtPcondsUAS}. More specifically,
let $\Theta$ be taken such that $\alpha(\mg)\neq 0$ on $\mgspace$, ${\Theta}'(\mg_{\ini})/\alpha(\mg_{\ini})\le0$ and ${\Theta}'(\mg_{\fin})/\alpha(\mg_{\fin})\le0$, as well as $\Theta(\mg_e)=0$ and ${\Theta}'(\mg_e)/\alpha(\mg_e)>0$ for some $\mg_e\in\interior{(\mgspace)}$. Then Condition \eqref{eq:C1cond} corresponds to the existence of a separatrix connecting $\mg_{\ini}$ and $\mg_{\fin}$, for which the corresponding function $\nvel:\mgspace\to\Rip$ can be found from \eqref{eq:IntOfABG}. We  utilize this procedure in the following example. 
\subsection{Simulation example: The ``butterfly'' robot }\label{sec:SimRes}
\begin{table}[t]
    \caption{Parameter values of the ``butterfly'' robot (BR).
    }
    \label{tab:SystemParams}{\tiny
    \begin{tabular}{ c c c c c }
    \toprule
       $m$ [\SI{}{\kilo\gram}] & $r_b$ [\SI{}{\meter}] & $J_b$ [\SI{}{\kilo\gram\meter\squared}]  & $J_f$ [\SI{}{\kilo\gram \meter\squared}] & $g$ [\SI{}{\meter\per\second\squared}]                   \\
          \midrule
         $3.0\times 10^{-3}$ & $1.09\times 10^{-2}$ &  $5.8\times 10^{-7}$ &  $8.9\times 10^{-4}$ & 9.81 
         \\ \bottomrule
    \end{tabular}
    }
\end{table}
Consider the ``butterfly'' robot (BR) illustrated in Figure~\ref{fig:BRmain}. Its shape is described by \eqref{eq:BRframe} with $a=1.14\times 10^{-1}$  and   $b=3.9\times 10^{-2}$, while the values of the
system parameters are given in Table~\ref{tab:SystemParams}. The  task we will consider is to maneuver the ball from $\varphi_{\ini}=\SI{0}{\radian}$ to $\varphi_{\fin}=\SI{2}{\radian}$.

\begin{figure}
    \centering
    \includegraphics[width=1.0\linewidth]{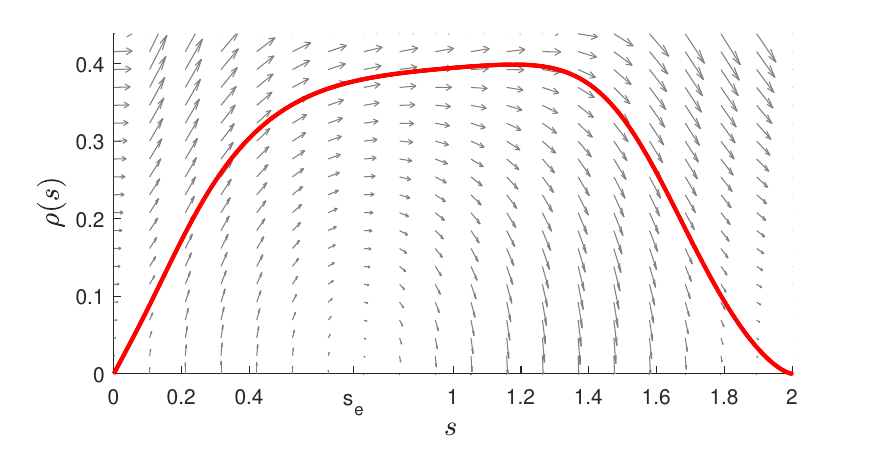}
    \caption{Phase portrait of \eqref{eq:reducedDynamics}, with the red curve the  solution of \eqref{eq:ABG}  satisfying  \textbf{P3} in Definition~\ref{def:s-param_mamever}.}
    \label{fig:rho}
\end{figure}

\begin{figure}
    \centering
    \includegraphics[width=0.95\linewidth]{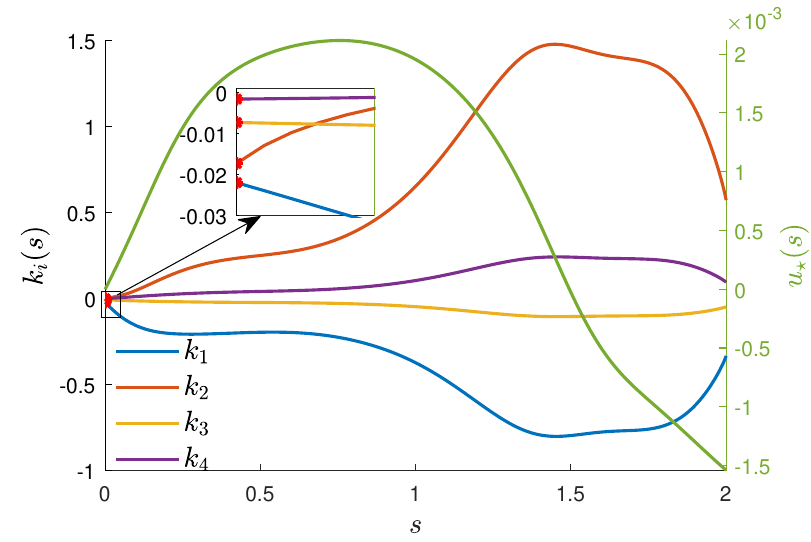}
    \caption{Found elements of $\lfb(\mg)=[k_1(\mg),k_2(\mg),k_3(\mg),k_4(\mg)]$ (left axis) and the nominal control input $\ac_\nom(\mg)$ (right axis).} 
    \label{fig:contGains}
\end{figure}

\textbf{Motion planning.} In light of Proposition~\ref{prop:BRprop},  consider the synchronization functions \eqref{eq:BRvc}
 with $\Theta(\mg)=k(\mg-\mg_{\ini})(\mg-\mg_e)(\mg_{\fin}-\mg)^2$,
where $\mg_{\ini}=0$,  $\mg_e\approx0.707$, $\mg_{\fin}=2$, and $k=0.01$. The corresponding unique (positive) solution to  \eqref{eq:ABG},  found using \eqref{eq:IntOfABG} and satisfying Property \textbf{P3} in Definition~\ref{def:s-param_mamever},  is shown in red  in Figure~\ref{fig:rho}. The corresponding nominal control input found from \eqref{eq:nomCont} can be seen in Figure~\ref{fig:contGains}, where it is measured relative to the right vertical axis.

\textbf{Projection operator.} We took $\Lambda=\text{diag}(0,1,0,0)$ in \eqref{eq:GenProjOp} with $\mgspace:=[\mg_{\ini},\mg_{\fin}]$, which is  equivalent to
$\prj(\state)=\sat_{s_{\ini}}^{s_{\fin}}(\varphi)=\max(\mg_{\ini},\min(\varphi,\mg_{\fin}))$.

\textbf{Control design.} 
Since the Jacobian linearization is  linearly controllable at both $\state_{\ini}=\state_\nom(\mg_{\ini})$ and $\state_{\fin}=\state_\nom(\mg_{\fin})$, we computed a pair of constant LQR-based feedback matrices $K_{\ini},K_{\fin}\in\Ri{m\times n}$  by solving the algebraic Riccati equations \eqref{eq:AREs} using the \texttt{CARE} command in MATLAB,  with $\Gamma_{\ini}=\Gamma_{\fin}=10^5$ and $Q_{\ini}=Q_{\fin}=\I{4}$. Note that the magnitude of $\Gamma_{\ini}$ and $\Gamma_{\fin}$ here simply reflects the small parameter values (see Table~\ref{tab:SystemParams}). 
 We then took $\lambda=0.5$, and formulated a semidefinite programming (SDP) problem  following Proposition~\ref{prop:LMIcontDesign} with the equality constraints \eqref{eq:EqContConstraints}. In order to discretize the differential LMI \eqref{eq:DLMI} into a finite number of LMIs, we took the elements of  the matrix functions $W$ and $Y$  as sixth-order Beziér polynomials, and  took \eqref{eq:DLMI} evaluated at 200 evenly spaced points as LMI constraints in the SDP. The resulting SDP  was then solved using the  YALMIP toolbox for MATLAB  \citep{lofberg2004yalmip}  together with the SDPT3 solver \citep{tutuncu2003solving}. Figure~\ref{fig:contGains} shows the  elements of the obtained $\lfb(\mg)=Y(\mg)W\inv(\mg)\in\Ri{1\times 4}$.

\textbf{Implementation.}
Following the discussion of Remark~\ref{remark:epsilonProjOp}, the projection operator was implemented as $\prj(\state)=\text{sat}_{s_{\ini}+\epsilon}^{s_{\fin}}(\varphi)$, where the dynamic variable $\epsilon\in[0,\epsilon_M]$ was governed by $\dot{\epsilon}=\epsilon_M\text{sign}\big(\epsilon_M-\|\state-\state_{\ini}\|\big)$ with $\epsilon_M=10^{-3}$ (similar results were obtained with a constant $\epsilon= \epsilon_M$). Since exact measurements of all the states were assumed to be given, the implementation of the controller \eqref{eq:MRfeedback} is straightforward: \textbf{Step 1:} Given $\state$, compute $\prj=\prj(\state)$; \textbf{Step 2:} Compute $\ac_\nom(\prj)$, $\lfb(\prj)$ and $\state_\nom(\prj)$ (e.g. using splines or lookup tables); \textbf{Step 3:} Take $\ac=\ac_\nom(\prj)+\lfb(\prj)\tvc$ with $\tvc=\state-\state_\nom(\prj)$. 

\begin{figure}
    \centering
    \includegraphics[width=1\linewidth]{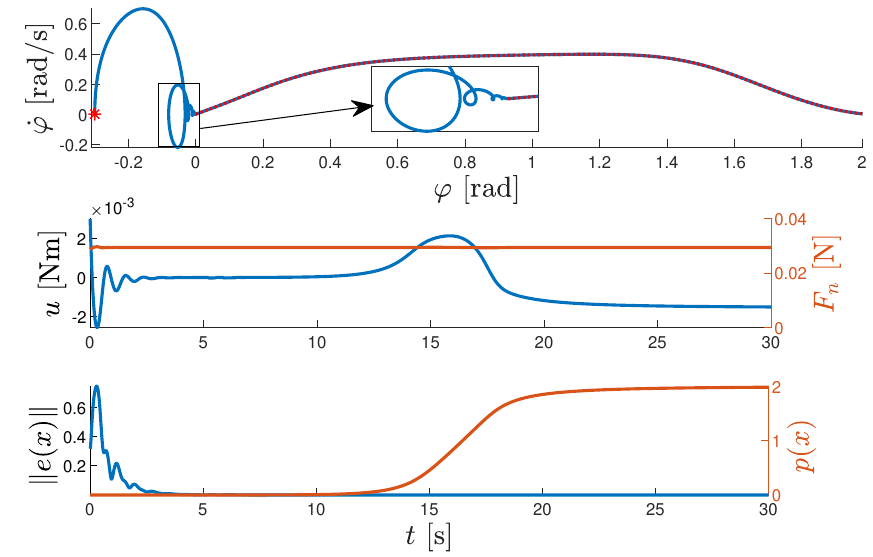}
    \caption{Response of the BR system  initialized close to $x_{\ini}$.}
    \label{fig:per_x0}
\end{figure}

\begin{figure*}
    \centering
    \includegraphics[trim={0.9cm 0cm 0.5cm 0.8cm},clip,width=1\linewidth]{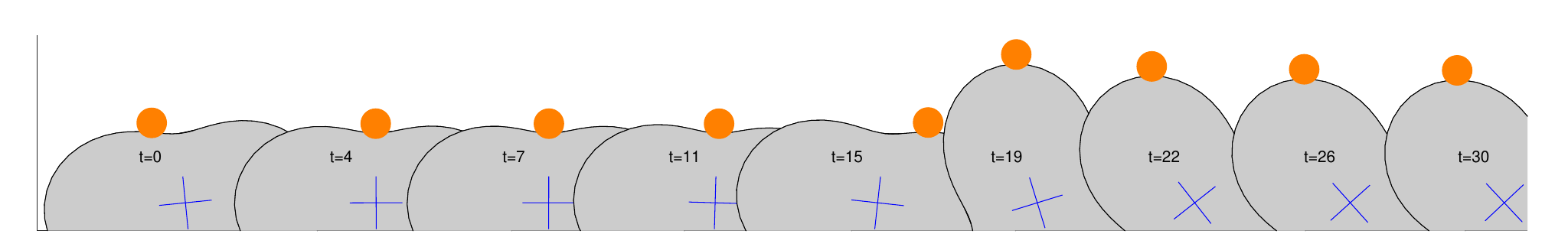}
    \caption{Snapshots of the configuration of the ``butterfly'' robot system corresponding to the response shown in Fig.~\ref{fig:per_x0}.}
    \label{fig:stopmotion}
\end{figure*}

\textbf{Simulation results.} The response of the system when starting with the initial  conditions $\state(0)=\state_{\ini}+\colvec(0.1,-0.3,0,0)$ is shown in Figure~\ref{fig:per_x0}, with some snapshots of the system's configuration shown in Figure~\ref{fig:stopmotion}.  
As the states are initially within the half-ball corresponding to $\state_{\ini}$, it can be seen that the controller  first brings the states close to $\state_{\ini}$, after which they then follow the nominal orbit to $\state_{\fin}$. Notice also that  Assumption \textbf{A2} holds, as the normal force $F_n$ between the ball and the frame is everywhere positive.

\begin{figure}
    \centering
    \includegraphics[width=1\linewidth]{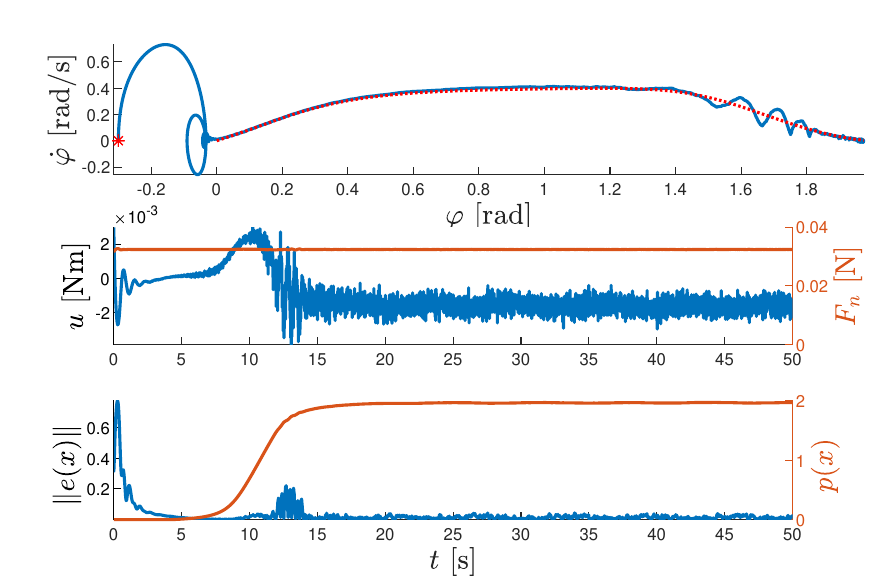}
    \caption{Response of the BR system initialized close to $x_{\ini}$, with white noise added to all the state measurements, with the mass of the ball increased by $10\%$  and subject to a small matched disturbance.}
    \label{fig:per_x0_dist}
\end{figure}

To test the sensitivity of the closed-loop system to noise and perturbations, we simulated the system with the same initial conditions, but with a small amount of white noise added to the measurements passed to the controller, with the actual mass of the ball, $m_b$, being $10\%$ larger than that assumed, as well as with the matched disturbance $10^{-4}\sin(t)$ added to the right-hand side of \eqref{eq:Manipulator equation}. The resulting system response is shown in Figure~\ref{fig:per_x0_dist}. 

\begin{figure}
    \centering
    \includegraphics[width=1\linewidth]{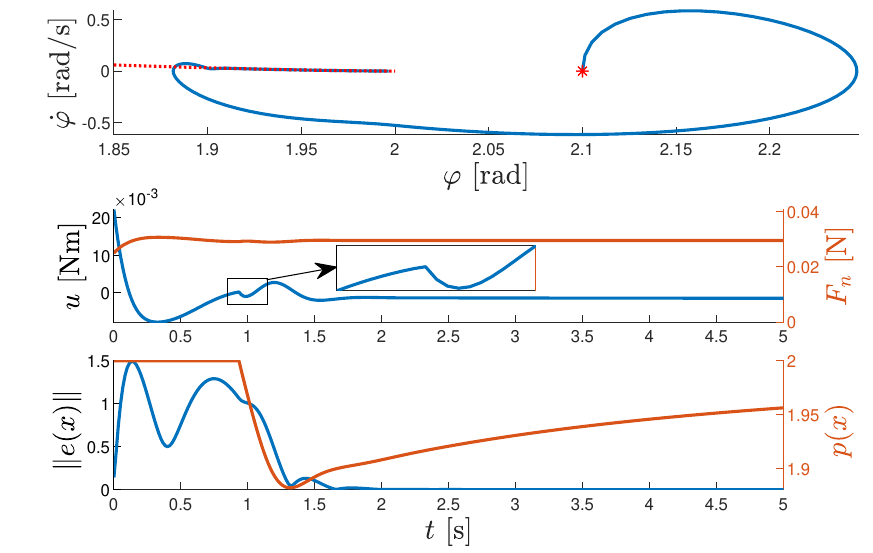}
    \caption{Response of the BR system initialized close to $x_{\fin}$.}
    \label{fig:per_xf}
\end{figure}

Figure~\ref{fig:per_xf} shows the system response for  $\state(0)=\state_{\fin}+\colvec(0.1,0.1,0,0)$. Interestingly, these initial conditions do not lie in the region of attraction of the linear feedback $\ac=\ac_\nom(\mg_{\fin})+\lfb(\mg_{\fin})(\state-\state_{\fin})$. Notice also that $\varphi$ becomes less than \SI{2}{\radian} just before $t= \SI{1}{\second}$, at which the gradient of the projection operator has a discontinuity. It can be seen that the smoothness of the control signal is violated at this time instant, but it is clear from the highlighted rectangle that Lipschitz continuity is still preserved. 

 \section{Discussion }\label{Sec:discussion}
\textbf{Is this Orbital Stabilization?} The main focus of this paper has been upon the stabilization of the set $\nomorb$ (see \eqref{eq:nomOrb}) corresponding to an assumed-to-be-known maneuver $\maneuver$. Even though this set consists of a heteroclinic orbit and its limit points, it may not be immediately clear that this form of set-stabilizing feedback can be referred to as an \emph{orbitally stabilizing} feedback. We however believe such a classification is not only justified, but that it is in fact an important one to make. To illustrate this point, consider the \emph{orbital stabilization problem} (see Prob.~\ref{probl:EOSofPPm}). As previously stated, it is equivalent to ensuring the asymptotic orbital stability \citep{hahn1967stability,leonov1995local,urabe1967nonlinear,zubov1999theory} of the desired motion.   
It therefore incorporates the  problem of stabilizing several important behaviors, including those corresponding to equilibria (trivial orbits),  limit cycles (periodic orbits) and PtP maneuvers (heteroclinic orbits). This motivates  developing general-purpose methods which can be used to control and stabilize these types of maneuvers (and more). Take, for instance, the method we have proposed in this paper: In the case of trivial orbits,  Theorem~\ref{theorem:MR} and Proposition~\ref{prop:LMIcontDesign} condenses down to a standard linear  feedback stabilizing the Jacobian linearization and to the satisfaction of an algebraic Lyapunov equation; whereas for nontrivial periodic orbits, a  control law of the form \eqref{eq:MRfeedback} satisfying \eqref{eq:PLDE}, e.g. found by solving the then periodic differential LMI \eqref{eq:DLMI}, will exponentially stabilize the desired orbit. 

\textbf{Rate of convergence.} 
A major (practical) limitation of the proposed scheme is the slow convergence away from the initial equilibrium.  In light of this issue, a possible ad hoc modification was proposed in Remark~\ref{remark:epsilonProjOp} as to ensure that the state do not remain too long about $\state_{\ini}$. The suggested modifications were, roughly speaking, based on removing the initial equilibrium  and instead starting part way along the maneuver, either by removing it altogether (static approach) or gradually moving away from it (dynamic approach). 
As an alternative way of handling this issue, especially the slow convergence away from the initial equilibrium, one can instead consider maneuvers where  $\state_{\ini}$ is  finite-time repellent with respect to $\nomorb$. If also $\state_{\fin}$ is finite-time attractive, then we refer to it as a \emph{finite-time} PtP maneuver. For such a maneuver, $\nvel(\cdot)$ can of course no longer be Lipschitz about $\mg_{\ini}$ and/or $\mg_{\fin}$. For instance, taking $\nvel(\mg)=\kappa|\mg-q_{\ini}|^{n_{\ini}} |q_{\fin}-\mg|^{n_{\fin}}$ for any $n_{\ini},n_{\fin}\in(0.5,1)$  in Example~\ref{example:DI} corresponds to a finite-time maneuver. Note, however, that for $\prj(\state)=\sat_{q_{\ini}}^{q_{\fin}}(q)$, the orbitally stabilizing feedback then cannot be Lipschitz about $\state_{\fin}$, as  $\nvel'(\mg)\to-\infty$ when $\mg\to\mg_{\fin}$.   Note also that for such a maneuver to exist in the solution space of an underactuated mechanical system,  the reduced dynamics  \eqref{eq:reducedDynamics} must a have certain type of singular point at the respective boundaries. Take, for example, 
 $\mg\Ddot{\mg}+(1-a)\dot{\mg}^2-b\mg(\mg-c)=0$ with  $a>3/2$ and $b,c>0$. It has a heteroclinic orbit connecting $\mg_{\ini}=0$ and $\mg_{\fin}=c$. Here $\mg_{\ini}$ is not only an equilibrium point, but also a singular point  of the type considered in \cite{surov2018new}, making it finite-time repellent with respect to the  orbit.

\section{Conclusion}\label{sec:conclusion}
We have introduced a method for inducing, via locally Lipschitz-continuous static state-feedback control, an asymptotically stable heteroclinic orbit in a nonlinear control system. Our suggested approach used a particular  parameterization of a known point-to-point maneuver, together with a so-called projection operator, as to merge a Jacobian linearization with a transverse linearization for the purpose of control design. Moreover, a possible way of constructing such a feedback by solving a semidefinite programming problem was suggested, while statements which may be used to plan such maneuvers for mechanical systems with one degree of underactuation using synchronization functions were provided. 

It was demonstrated that the approach could be used to solve the challenging nonprehensile manipulation problem of rolling a ball, in a stable manner, between any two points upon a smooth actuated planer frame. This provided a general solution applicable to a number of well-known nonlinear systems, including the ball-and-beam, the disk-on-disk and the ``butterfly'' robot. The approach was successfully demonstrated on the latter system in numerical simulations. 

\begin{ack}                     
The authors are grateful to the anonymous reviewers; their insightful comments and suggestions have  helped us significantly improve the quality of this work.
\end{ack}

\bibliographystyle{abbrvnat}
\bibliography{references}       

\appendix
\section{Appendix}

\subsection{Proof of Proposition~\ref{prop:GenProjOp}}\label{app:GenProjOpProof}
We need to show that all the conditions in Definition~\ref{def:ProjOp}  are satisfied. To this end, we begin by differentiating the terms inside the brackets in \eqref{eq:GenProjOp} with respect to $\mg$, from which we obtain the  function
\begin{equation*}
    z(\state,\mg):= (\state-\state_\nom(\mg))\transp\left[\Lambda'(\mg)(\state-\state_\nom(\mg))-2\Lambda(\mg)\nomflow(\mg)\right].
\end{equation*}
Since $\frac{\partial}{\partial \mg}z(\state,\mg)\rvert_{\state=\state_\nom(\mg)}=2\nomflow\transp(\mg)\Lambda(\mg)\nomflow(\mg)>0$ and  $z(\state_\nom(\mg),\mg)\equiv 0$, Condition \textbf{C1} is implied.
Moreover, by noting from Property \textbf{P1} in Definition~\ref{def:s-param_mamever} that the curve $\state_\nom(\cdot)$  has bounded curvature  and  is not self-intersecting,  the implicit function theorem  \cite[Thm. 3.1.10]{berger1977nonlinearity} ensures that there  exists, in a certain vicinity of each point on $\nomorb$, a unique function $\prj(\state)$  satisfying $z(\state,\prj(\state))\equiv 0$, which  in turn implies that $\prj(\state)$ solves   \eqref{eq:GenProjOp}. Thus, for  $\prjspace\subset\Ri{n}$ a sufficiently small neighborhood of $\nomorb$, the requirement $\mathfrak{L}(\state_\nom(\prj(\state)),\state)\subset\mathcal{B}_\epsilon(\state_\nom(\mg))\subset\prjspace$ ensures the uniqueness of a solution to \eqref{eq:GenProjOp} within $\tubespace:=\{\state\in\prjspace: \ z(\state,\prj(\state))=0\}$. Moreover, if $\state\in\tubespace$, then   \cite[Cor. 3.1.11]{berger1977nonlinearity}
\begin{equation*}
    \jac{\prj}(\state)=\frac{\nomflow\transp \Lambda-\tvc\transp \Lambda'}{\nomflow\transp \Lambda\nomflow+\tvc\transp\left[\frac{1}{2}\Lambda''\tvc-2\Lambda'\nomflow-\Lambda\nomflow' \right]}
\end{equation*}
 for such a solution $\prj=\prj(\state)$,
with  $\tvc:=\state-\state_\nom(\prj)$, and where we have omitted the $\prj$-arguments to shorten the notation, i.e. $\nomflow=\nomflow(\prj)$ etc. Hence $\jac{\prj}(\cdot)$ is nonzero and $\cont{r}$ (as $\Lambda\nomflow'$ is)  within  $\tubespace\subset\Ri{n}$, with $\projjac(\mg):=\jac{\prj}(\state_\nom(\mg))$   given by \eqref{eq:genProjOpJac} therein.

 What remains is therefore to show the  parts of \textbf{C2} and \textbf{C3} in Definition~\ref{def:ProjOp} relating to the sets $\mathcal{H}_{\ini}$ and $\mathcal{H}_{\fin}$ also hold. Let us assume these sets exist. 
 Due to the expression for $\jac{\prj}(\cdot)$ above, which is valid within $\tubespace$, together with $\projjac \nomflow=1$, it follows that sufficiently close to $\state_{\fin}$ the states will leave $\tubespace$ and enter $\mathcal{H}_\fin$ if they go in the direction $\nomflow(\mg_{\fin})$ when on  $\prjspace_{\fin}:=\closure(\tubespace)\cap\closure(\mathcal{H}_{\fin})$.
 Take $\prjspace$ such that any $\state\in\mathcal{H}_\fin$ can be written as $\state=\chi_\fin+c\nomflow(\mg_\fin)$ for some  $\chi_\fin\in\prjspace_\fin$ and $c>0$. Since $z(\chi_\fin,\mg_\fin)=0$, one can, for $\lVert \state-\state_\fin\rVert$ sufficiently small, always find a Lagrange multiplier $\mu_\fin>0$ associated with the inequality constraint $\mg_\fin-\mg\ge 0$, such that $z(\state,\mg_\fin)+\mu_\fin=0$. Thus $\mg_\fin$ is a minimizer  by the Karush--Kuhn--Tucker  conditions.
 Moreover, due to the constraint $\mathfrak{L}(\state_\nom(\mg),\state)\subset\mathcal{B}_\epsilon(\state_\nom(\mg))\subset\prjspace$ and the condition $\nomflow\transp \Lambda\nomflow >0$, we can always take both
 $\prjspace$ and $\mathcal{H}_\fin$ to be sufficiently small as to guarantee that $\mg_\fin$ is the unique minimizer of \eqref{eq:GenProjOp} for all $\state\in\mathcal{H}_\fin$.
 Using the same arguments about $\prjspace_{\ini}:=\closure(\mathcal{H}_{\ini})\cap\closure(\tubespace)$, the existence of $\mathcal{H}_{\ini}$ and $\mathcal{H}_{\fin}$ in Condition \textbf{C2} is therefore implied, and the requirements of \textbf{C3} are met.

\subsection{Proof of Lemma~\ref{lemma:altFuncRep}}
\label{app:altFuncRepProof}
According to Taylor's theorem  (see, e.g., \cite[Thm. 2.1.33]{berger1977nonlinearity}),   $\sigma(\state)=\jac{\sigma}(y)(\state-y)+\bigO(\|\state-y\|^2)$ holds
for all $\state$ in some neighborhood of  a fixed $y\in\nomorb$. 
Due to the properties of a projection operator (see Def.~\ref{def:ProjOp}), there is a neighborhood
$\prjspacesub\subseteq\prjspace$ of $\nomorb$, such that  $\mathfrak{L}(\xp(\state),\state)\subset\prjspacesub$ for all $\state\in\prjspacesub$.
Hence, for any $\state\in\prjspacesub$, we may take $y=\xp(\state)$  to obtain \eqref{eq:TErewriteFncs}.
 Due to  $\prj(\cdot)$ being at least   $\cont{1}$ within  $\mathcal{H}_{\ini}$, $\tubespace$ and $\mathcal{H}_{\fin}$,  the validity of  \eqref{eq:TErewriteFncs} is ensured almost everywhere within $\prjspacesub$.
 
\subsection{Proof of Proposition~\ref{prop:dynamicsOfCoords}}\label{app:dynamicsOfCoordsProof}
Recall  that $\jac{\tvc}(\state)=\I{n}$ whenever $\state$ is within either $\mathcal{H}_{\ini}$ or $\mathcal{H}_{\fin}$. By  computing the Jacobian matrix of the right-hand side of \eqref{eq:errorDyn} and  using \eqref{eq:TErewriteFncs}, we therefore readily obtain \eqref{eq:JacDyn}. 
In order to also show that \eqref{eq:TubeDyn}
 is valid within $\tubespace$, we note that \eqref{eq:TErewriteFncs} must also be valid for the  function $\tvc(\cdot)$ itself within the interior of  $\tubespace$, as $\prj\in\cont{2}$ therein. Let  $\prj=\prj(\state)$ and recall that   $\tvcjac^2=\tvcjac$ (see Lem.~\ref{lemma:tvcjac}). Applying \eqref{eq:TErewriteFncs} to each element of $\tvc$, and then multiplying from the left by $\tvcjac(\prj)$,  one finds that
\begin{equation}
\label{eq:TCTaylorExp}
    \tvc(\state)=\tvcjac(\prj)\tvc(\state)+\nomflow(\prj)l(\state)
\end{equation}
must hold for  $\state\in\tubespace$, with $l:\Ri{n}\to\Ri{}$ some $\cont{2}$ function   satisfying  $\|l(\state)\|=\bigO(\|\tvc\|^2)$. 
 Using the fact that  $\jac{\tvc}(\state)=\I{n}-\nomflow(\prj(\state))\jac{\prj}(\state)$ whenever $\state\in{\tubespace}$, the Jacobian matrix of the right-hand side of \eqref{eq:errorDyn} can also be computed inside $\tubespace$. By  writing it in the form  \eqref{eq:TErewriteFncs} and using \eqref{eq:TCTaylorExp}, one obtains \eqref{eq:TubeDyn}.
 
The above still applies even if there  are points such that $\mathfrak{L}(\xp(\state),\state)$ does not remain in a given subset of $\prjspace$, regardless of how small $\mathcal{N}(\nomorb)$ is taken. Indeed, within $\mathcal{H}_i$ one can use the equivalence between the right-hand side of \eqref{eq:errorDyn} with the function  obtained  by fixing $\prj=\mg_i$. Moreover, Property \textbf{P1} in Definition~\ref{def:s-param_mamever} ensures that one can always find a function which is $\cont{2}$-smooth in $\neighb(\nomorb)$ and equivalent to the right-hand side of \eqref{eq:errorDyn} for all $\state$ in $\neighb(\nomorb)\cap\tubespace$. Specifically, there exists an $\epsilon>0$ such that one can extend the maneuver at its boundaries in the appropriate direction along $\nomflow(\mg_i)$ and $\ac_s'(\mg_i)$ for $\lvert\mg-\mg_i\rvert<\epsilon$. An appropriate projection onto this extended maneuver,  which is equivalent to $\prj$ in $\tubespace$ and which is $\cont{2}$-smooth in the whole of $\neighb(\nomorb)$, can then be constructed and used to define the aforementioned function.

\subsection{Proof of Proposition~\ref{Prop:Uniqueness}}\label{app:uniquenessProp}
In the following, we will sometimes omit the $\mg$-arguments as to shorten the notation.
 Given a solution $R(\mg)$ to \eqref{eq:PLDE}, let $R_\perp:=\tvcjac\transp R  \tvcjac$. Clearly $R_\perp=\tvcjac\transp R_\perp \tvcjac$ then holds by Lemma~\ref{lemma:tvcjac}. 
    Differentiating  $R_\perp(\mg)$ with respect to $\mg$ yields ${R}_\perp'=\Big(\frac{d}{d\mg}{\tvcjac\transp}\Big) R \tvcjac +{\tvcjac\transp} {R}' \tvcjac +{\tvcjac\transp} R \Big(\frac{d}{d\mg}{\tvcjac}\Big)$.
By then using that $\big( \frac{d}{d\mg}{\tvcjac}\big) \tvcjac=-\nomflow\nomflow\transp\hess{\prj}(\state_\nom)$, one finds, by inserting the above expression for ${R}_\perp'$ into \eqref{eq:MLDE}, that \eqref{eq:MLDE} holds if   $R(\mg)$  satisfies \eqref{eq:PLDE}. 

To show that the converse holds as well, let $R_\perp(\mg)=\tvcjac\transp(\mg)R_\perp(\mg)\tvcjac(\mg)$ solve \eqref{eq:PLDE}. Taking    then ${R}(\mg):={R}_\perp(\mg)+h_R(\mg)\projjac\transp(\mg)\projjac(\mg)$, with 
 $h_{R}:\mgspace\to\Ri{}_{>0}$  an arbitrary smooth function, one can easily show, using the properties stated in Lemma~\ref{lemma:tvcjac}, that  $R(\mg)$ satisfies \eqref{eq:PLDE}.

What remains is therefore to show that a solution $R_\perp(\mg)=\tvcjac\transp(\mg)R_\perp(\mg)\tvcjac(\mg)$ to \eqref{eq:PLDE} is unique. In this regard, first note that by Lemma~\ref{lemma:tvcjac} and the relation $\projjac\nomflow\equiv 1$, we can always find some   $\omega\transp,\mathcal{J}:\mgspace\to\Ri{n\times n-1}$ which are sufficiently smooth and satisfy $ \omega(\mg)\nomflow({\mg})\equiv \0{n-1\times 1}$, $\projjac({\mg})\mathcal{J}(\mg)\equiv \0{1\times n-1}$ and $ \omega(\mg)\mathcal{J}(\mg)\equiv \I{n-1}$
 for all $\mg\in\mgspace$. In particular, we will here take $\mathcal{J}$ satisfying $\dot{\mathcal{J}}=-\nomflow \dot{\projjac}\mathcal{J}$. 
 This allows us to  write $\tvcjac({\mg})=\Omega(\mg)E\Omega\inv(\mg)$ in which $E:=\text{diag}(0,\I{n-1})$, $\Omega(\mg):=\big[\nomflow({\mg}), \mathcal{J}(\mg)\big]$ and $\Omega\inv(\mg)=\big[\projjac\transp({\mg}), \omega\transp(\mg) \big]\transp$.
We can then equivalently rewrite \eqref{eq:PLDE} as
  \begin{align*}
      E\Omega\transp\big[\dot{R}+A_{cl}\transp (\Omega\inv)\transp &E\Omega\transp \hat{R}_\perp+\hat{R}_\perp \Omega E\Omega\inv A_{cl}
      \\
      &-\dot{\projjac}\transp \nomflow\transp R-R\nomflow\dot{\projjac}+Q\big]\Omega E= \0{n},
 \end{align*}
 with $\dot{\projjac}=\nomflow\transp \hess{\prj}(\state_\nom)\nvel$. It can further be shown that the parts of this equation which are not trivially zero correspond to the following matrix differential equation:
 \begin{equation*}
     \mathcal{A}\transp \mathcal{R}_\perp+\mathcal{R}_\perp\mathcal{A}+\mathcal{J}\transp\big[\dot{R}-\dot{\projjac}\transp \nomflow\transp R-R\nomflow\dot{\projjac}\big]\mathcal{J}+\mathcal{Q}_\perp =\0{n},
 \end{equation*}
 where $\mathcal{A}(\mg):=\omega(\mg)A_{cl}(\mg)\mathcal{J}(\mg)$, while the matrix functions $\mathcal{R}_\perp(\mg):=\mathcal{J}\transp(\mg) R(\mg) \mathcal{J}(\mg)$ and  $\mathcal{Q}_\perp(\mg):=\mathcal{J}\transp(\mg)Q(\mg)  \mathcal{J}(\mg)$
 evidently are both  $\cont{1}$-smooth, symmetric and positive definite. Since $\dot{\mathcal{J}}=-\nomflow \dot{\projjac}\mathcal{J}$, we have $ \dot{\mathcal{R}}_\perp=\mathcal{J}\transp\left[\dot{R}-\dot{\projjac}\transp \nomflow\transp R-R\nomflow\dot{\projjac}\right]\mathcal{J}$.
  We can therefore rewrite the above equation as
 \begin{equation}\label{eq:PLDEredfull}
     {\mathcal{R}}_\perp'(\mg)\nvel(\mg)=-\mathcal{A}\transp(\mg) \mathcal{R}_\perp(\mg)-\mathcal{R}_\perp(\mg)\mathcal{A}(\mg)-\mathcal{Q}_\perp(\mg).
 \end{equation}
 In order to show uniqueness, we use hypotheses that $\nvel(\mg_{\ini})=0$ and $\nvel'(\mg_{\ini})>0$.
 Hence, due to both $\mathcal{R}_\perp(\mg_{\ini})$ and $\mathcal{Q}_\perp(\mg_{\ini})$ being members of $\Mats_{\succ 0}^{n-1}$ and satisfying the algebraic   Lyapunov equation \eqref{eq:PLDEredfull} for $\mg =\mg_{\ini}$, it follows that the matrix $\mathcal{A}(\mg_{\ini}):=\omega(\mg_{\ini})A_{cl}(\mg_{\ini})\mathcal{J}(\mg_{\ini})$ must necessarily be Hurwitz, which  in turn implies that $\mathcal{R}_\perp(\mg_{\ini})$ is unique 
\cite[Theorem 4.6]{khalil2002nonlinear}.  
Since the right-hand side of \eqref{eq:PLDEredfull} is continuously differentiable, it then has a unique solution $\mathcal{R}_\perp(\mg)$ satisfying $\mathcal{R}_\perp(\mg(t_{\ini}))=\mathcal{R}_\perp(\mg_{\ini})$.
Consequently $\hat{R}_\perp(\mg)=\tvcjac\transp(\mg)R(\mg)\tvcjac(\mg)=\omega\transp(\mg) \mathcal{J}\transp (\mg)R(\mg) \mathcal{J}(\mg) \omega(\mg)=\omega\transp(\mg)\mathcal{R}_\perp(\mg) \omega(\mg)$ is also unique.
This concludes the proof.

\subsection{Proof of Theorem~\ref{theorem:MR}}\label{app:MainReulstProof}
    By the reduction principle stated in Corollary 11 in \cite{el2013reduction}, $b)\implies a)$. Moreover, the forward invariance property stated in $b)$ holds due to the existence of the maneuver (see Property \textbf{P5} in Def.~\ref{def:s-param_mamever}) and the properties of projection operators (see Def.~\ref{def:ProjOp}). 
    We therefore claim that $c)\implies b)$.
    Indeed,  first note that $V$ is  differentiable everywhere in $\prjspace$ except at the hypersurfaces (having zero  Lebesgue measure; cf. Rademacher's theorem \citep{clarke2008nonsmooth}) $\prjspace_{\ini}:=\lim_{\mg\to\mg_{\ini}^+}\Pi(\mg)$ and $\prjspace_{\fin}:=\lim_{\mg\to\mg_{\fin}^-}\Pi(\mg)$  (see \eqref{eq:MPS} for the definition of $\Pi$).  
        Denote $v(t)=V(\state(t))$ and consider the upper-right (Dini) derivative  $v^+(t)$ of $v(t)$, defined by $v^+(t):=\limsup_{h\to 0^+}\frac{1}{h}\left[v(t+h)-v(t)\right]$. At  $\state=\state(t)$, this is equivalent to  (see \cite{yoshizawa1975stability})
    \begin{equation*}
        V^+(x)=\limsup_{h\to 0^+}\frac{1}{h}\left[V(\state +h\ffunc_{cl}(\state))-V(\state)\right].    
    \end{equation*}
    Here $\ffunc_{cl}(\state):=\ffunc(\state)+\gfunc(\state)\big[\ac_\nom(\prj)+\lfb(\prj)\tvc\big]$ corresponds to the right-hand side of the autonomous closed-loop system, which we recall is locally Lipschitz and thus guaranteeing (local) existence and uniqueness of solutions. It is known  \citep{clarke2008nonsmooth} that  the following holds:
    \begin{equation*}
        V^+(\state)\le \limsup_{y\to\state}\left\{DV(y)\ffunc_{cl}(x): \ y\notin \prjspace_{\alpha}\cup \prjspace_{\omega}\right\}.
    \end{equation*}
Hence $c)$ implies  $v^+(t)\le-\mu \cdot v(t)$ holding for all $t\ge t_0$ if the system is initialized within some  neighborhood $\mathfrak{T}$  at time $t_0$. Thus $c)\implies b)$ follows from the comparison lemma (see, e.g., \cite{yoshizawa1975stability,khalil2002nonlinear}).      
    
    What remains is therefore to show that the theorem's hypotheses  imply $c)$.
    Under the assumption that $V$ is differentiable at some $x$ in $\prjspace$, one finds, using  the shorthand notation $\prj=\prj(\state)$, that its time derivative is 
\begin{equation}\label{eq:LyapFuncDer}
    \dot{V}= \dot{\tvc}\transp R(\prj)\tvc+\tvc\transp\left[R'(\prj)\jac{\prj}(\state)\dot{\state}\right]\tvc+\tvc\transp R(\prj)\dot{\tvc}.
\end{equation}
Hence, whenever  $\state$ is within the interior of either of the sets $\interior{\mathcal{H}}_i$, $i\in\{\ini,\fin\}$, where $\|\jac{\prj}(\state)\|=0$,
one has by \eqref{eq:JacDyn} and the ALEs \eqref{eq:ALEs}
that the following holds therein:
\begin{equation}\label{eq:LyapDivH}
    \dot{V}=-\tvc\transp Q_i \tvc+\bigO(\|\tvc\|^3).
\end{equation}
Whenever $\state$ is in $\tubespace$, one instead has $\jac{\prj}(\state)\dot{\state}=\nvel(\prj(\state))+\bigO(\|\tvc\|)$ (this follows from  the first-order Taylor expansion about $\state_\nom(\prj(\state))$ and by using \eqref{eq:nvel}). Thus by \eqref{eq:TubeDyn}, \eqref{eq:TCTaylorExp}  and \eqref{eq:LyapFuncDer} we  obtain, for $ \state\in\tubespace $:
\begin{align*}
    \dot{V}=&\tvc\transp\tvcjac\transp\big[A_{cl}\transp\tvcjac\transp R+R\tvcjac A_{cl}
    \\ \nonumber &     
    +\nvel\big[R'-({\projjac'})\transp \nomflow\transp R-R\nomflow\projjac'\big]
    \Big]\tvcjac\tvc+\bigO(\|\tvc\|^3).
\end{align*}
 Since  a solution $R_\perp(\mg)$ to \eqref{eq:MLDE} implies a solution to \eqref{eq:PLDE} (see Prop.~\ref{Prop:Uniqueness}), we thus obtain, using also \eqref{eq:TCTaylorExp},   that
\begin{equation}\label{eq:LyapDivT}
    \dot{V}
    = -\tvc\transp Q_\perp(\prj)\tvc+\bigO(\|\tvc\|^3).
\end{equation}
Thus, by \eqref{eq:LyapDivH} and \eqref{eq:LyapDivT}, there exists some  constant ${\mu}>0$ such that the differential inequality $\dot{V}\le -{\mu} V$ holds almost everywhere (or everywhere if one considers $V^+(x)$)  within a neighborhood $\mathfrak{T}$ of $\nomorb$ where $\lVert\tvc\rVert$ is sufficiently small.
This concludes the proof. 
 
 \subsection{Proof of Proposition~\ref{prop:LMIcontDesign}}\label{app:proofOfDLMI}
 Let us first demonstrate that the ALEs \eqref{eq:ALEs} are satisfied. To this end, we  note that the constant matrix $A_{cl}(\mg_{\ini})=A_\mg(\mg_{\ini})+B_\mg(\mg_{\ini})Y(\mg_{\ini})W\inv(\mg_{\ini})$ is Hurwitz. Thus by Theorem 1 in \cite{bernussou1989linear},
 there exists a matrix $\hat{Q}_{\ini}\in\Mats_{\succ 0}^{n}$ such that $\text{sym}\left[A_\mg(\mg_{\ini})W(\mg_{\ini})+B_\mg(\mg_{\ini})Y(\mg_{\ini})\right]=-\hat{Q}_{\ini}$, where $\text{sym}[A]=A+A\transp$.
  For $R(\mg_{\ini})=W\inv(\mg_{\ini})$ we may therefore take $Q_{\ini}=W\inv(\mg_{\ini})\hat{Q}_{\ini}W\inv(\mg_{\ini})$ in  \eqref{eq:ALE0}. The exact same arguments can be used for the point $\mg_{\fin}$. 
  
Let us now show that a matrix function $W(\cdot)$ solving the differential LMI \eqref{eq:DLMI} is equivalent to a solution $R(\cdot)$ to \eqref{eq:PLDE} (and therefore also a solution $R_\perp$ to \eqref{eq:MLDE}). For this purpose, recall that for any smooth  nonsingular matrix function $W:\mgspace\to\Ri{n\times n}$ one has $\frac{d}{d\mg}W\inv(\mg)=-W\inv(\mg)[\frac{d}{d\mg}W(\mg)] W\inv(\mg)$. Thus taking $R(\mg):=W\inv(\mg)$ and dropping the $\mg$-argument to keep the notation short, we obtain the following from \eqref{eq:DLMI}:   
$\nvel R'\preceq-\text{sym}[RA_\perp+RB_\perp K
     +\lambda R\tvcjac]$.
Multiplying from the left by $\tvcjac\transp$ and by $\tvcjac$ from the right, this can be written as
$\tvcjac\transp \text{sym}\Big[R\tvcjac A_{cl}+\lambda R
    +\nvel\big(2\inv R'-R\nomflow\nomflow\transp\hess{\prj}(\state_\nom)\big) \Big]\tvcjac\preceq \0{n}$.
Hence, as $R=W\inv\in\Mats_{\succ 0}^n$ and $\lambda$ is strictly positive, there must exist a  $\cont{1}$-smooth matrix-valued function $Q_\perp:\mgspace \to\Mats_{\succ 0}^{n}$ such that $R$ solves the PrjLDE \eqref{eq:PLDE}.

\subsection{Proof of Theorem~\ref{theorem:PtPcondsUAS}}\label{app:PtPcondUASproof}
    The boundary conditions imposed on $\vc(\cdot)$ are obvious, whereas those on $\vc'(\cdot)$ are obtained directly from \eqref{eq:nonVanCondUndAcSys} by setting $\nvel(\mg_{\ini})=\nvel(\mg_{\fin})=0$. The condition $\alpha(\mg)\neq 0$  ensures  the uniqueness and smoothness of the solutions to \eqref{eq:reducedDynamics}. Moreover, it implies that the integrating factor \eqref{eq:PsiDef} is  both nonzero and bounded on $\mgspace$, such that \eqref{eq:ABG}$\iff$\eqref{eq:IntOfABG}  on $\mgspace$ by the fundamental  theorem of calculus.  By taking $\mg_1=\mg_\ini$ and $\mg_2=\mg$ in \eqref{eq:IntOfABG},  the function  $\nvel(\mg)=\sqrt{-2\frac{\Psi(\mg,\mg_{\ini})}{\alpha^2(\mg)}\int_{\mg_\ini}^{\mg}\Psi(\mg_\ini,\tau)\alpha(\tau)\gamma(\tau)d\tau}$ can be obtained. Clearly it is smooth, bounded and  satisfies \eqref{eq:ABG} on $\mgspace$, while $\nvel(\mg_\fin)=0$ due to \eqref{eq:C1cond}. To show that it is also real and strictly positive on $\interior{\mgspace}$, it suffices to note that the smooth function $\Upsilon(\mg):=\gamma(\mg)/\alpha(\mg)$ (which has the same sign as $\alpha(\mg)\gamma(\mg)$) is strictly negative on $(\mg_\ini,\mg_e)$ and strictly positive on $(\mg_e,\mg_\fin)$ as $\Upsilon(\bar{\mg})=0$ and $\Upsilon'(\bar{\mg})={\upsilon}(\bar{\mg})$ $\forall \bar{\mg}\in\{\mg_\ini,\mg_e,\mg_\fin\}$. Thus the term inside the square root is strictly positive on $(\mg_\ini,\mg_e)$.
     Since $\gamma(\mg)\neq 0$ for all $\interior{(\mgspace)}\backslash\{\mg_e\}$, and $\nu(\mg_e)>0$, the terms inside the square root, and therefore also the function $\nvel$,  must remain strictly positive on $(\mg_e,\mg_\fin)$.  
\end{document}